\documentclass[sigconf]{acmart}

\usepackage{booktabs,tabularx}
\usepackage{array}
\usepackage{graphicx}
\usepackage{caption}
\usepackage{subcaption}
\newcolumntype{Y}{>{\raggedright\arraybackslash}X}

\AtBeginDocument{%
  }

\setcopyright{acmlicensed}
\copyrightyear{2018}
\acmYear{2018}
\acmDOI{XXXXXXX.XXXXXXX}
\acmConference[Conference acronym 'XX]{Make sure to enter the correct
  conference title from your rights confirmation email}{June 03--05,
  2018}{Woodstock, NY}
\acmISBN{978-1-4503-XXXX-X/2018/06}

\usepackage[most]{tcolorbox}
\tcbuselibrary{breakable,skins}
\usepackage{tabularx,booktabs}
\usepackage{enumitem}
\usepackage{ragged2e}
\usepackage{xcolor}

\setlist[itemize]{topsep=2pt,itemsep=2pt,parsep=0pt,leftmargin=1.2em}
\setlist[enumerate]{topsep=2pt,itemsep=2pt,parsep=0pt,leftmargin=1.4em}

\definecolor{rubricbg}{RGB}{249,250,252}
\definecolor{rubricbd}{RGB}{210,214,220}
\definecolor{rubricht}{RGB}{33,37,41}

\newtcolorbox{RubricBox}[1][]{%
  enhanced,
  breakable,
  sharp corners,
  colback=rubricbg,
  colframe=rubricbd,
  coltitle=rubricht,
  fonttitle=\bfseries,
  title={#1},
  boxrule=0.6pt,
  left=10pt,right=10pt,top=10pt,bottom=10pt,
  attach boxed title to top left={yshift=-2pt,xshift=6pt},
  boxed title style={
    colback=white,
    colframe=rubricbd,
    boxrule=0.6pt,
    sharp corners,
    padding=3pt 6pt
  }
}

\newcolumntype{Y}{>{\RaggedRight\arraybackslash}X}

\begin{document}

\title{ClassMind: Scaling Classroom Observation and Instructional Feedback with Multimodal AI }

\author{Ao Qu}
\authornote{Equal contribution}
\affiliation{%
  \institution{Massachusetts Institute of Technology}
  \city{Cambridge}
  \state{Massachusetts}
  \country{United States}}
\email{qua@mit.edu}

\author{Yuxi Wen}
\authornotemark[1]
\affiliation{%
  \institution{Michigan State University}
  \city{East Lansing}
  \state{Michigan}
  \country{United States}}
\email{wenyuxi@msu.edu}

\author{Jiayi Zhang}
\authornotemark[1]
\affiliation{%
  \institution{National University of Singapore}
  \city{Singapore}
  \country{Singapore}}

\author{Yunge Wen}
\affiliation{%
  \institution{New York University}
  \city{New York}
  \state{New York}
  \country{United States}}

\author{Yibo Zhao}
\affiliation{%
  \institution{Massachusetts Institute of Technology}
  \city{Cambridge}
  \state{Massachusetts}
  \country{United States}}
\email{yzhao231@jh.edu}

\author{Alok Prakash}
\affiliation{%
  \institution{M3S, Singapore-MIT Alliance for Research and Technology}
  \city{Singapore}
  \country{Singapore}}
\email{alok.prakash@smart.mit.edu}

\author{Andrés F. Salazar-Gómez}
\affiliation{%
  \institution{MIT Open Learning, Massachusetts Institute of Technology}
  \city{Cambridge}
  \state{Massachusetts}
  \country{United States}}
\email{salacho@mit.edu}

\author{Paul Pu Liang}
\affiliation{%
  \institution{Massachusetts Institute of Technology}
  \city{Cambridge}
  \state{Massachusetts}
  \country{United States}}
\email{ppliang@mit.edu}

\author{Jinhua Zhao}
\affiliation{%
  \institution{Massachusetts Institute of Technology}
  \city{Cambridge}
  \state{Massachusetts}
  \country{United States}}
\email{jinhua@mit.edu}

\renewcommand{\shortauthors}{Ao et al.}

\begin{abstract}
classroom observation—one of the most effective methods for teacher development—remains limited due to high costs and a shortage of expert coaches. We present \textbf{ClassMind}, an AI-driven classroom observation system that integrates generative AI and multimodal learning to analyze classroom artifacts (e.g., class recordings) and deliver timely, personalized feedback aligned with pedagogical practices. At its core is AVA-Align, an agent framework that analyzes long classroom video recordings to generate temporally precise, best-practice-aligned feedback to support teacher reflection and improvement. Our three-phase study involved participatory co-design with educators, development of a full-stack system, and field testing with teachers at different stages of practice. Teachers highlighted the system’s usefulness, ease of use, and novelty, while also raising concerns about privacy and the role of human judgment, motivating deeper exploration of future human–AI coaching partnerships. This work illustrates how multimodal AI can scale expert coaching and advance teacher development.
\end{abstract}


\keywords{Do, Not, Use, This, Code, Put, the, Correct, Terms, for,
  Your, Paper}

\begin{teaserfigure}
  \includegraphics[width=\textwidth]{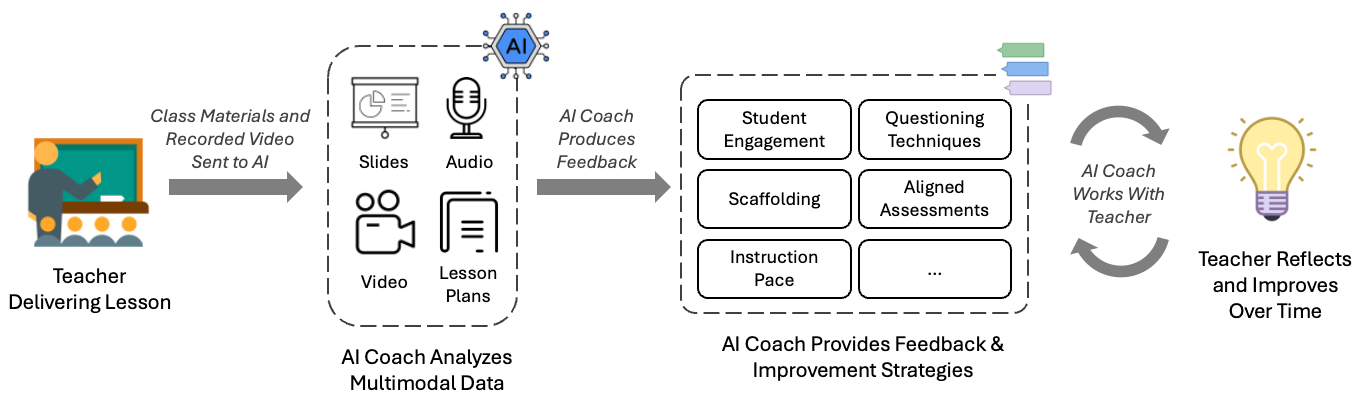}
  \caption{\textbf{ClassMind system pipeline.} Teachers upload classroom recordings and supporting materials (slides, audio, lesson plans, video). The AI coach performs multimodal analysis to integrate diverse inputs, applies long-context video understanding to process entire lessons, and generates rubric-aligned feedback across pedagogical dimensions (e.g., student engagement, questioning, scaffolding). Built on the AVA-Align framework for temporally precise, rubric-grounded classroom analysis, ClassMind is a novel open-source platform that delivers scalable AI feedback. Feedback is structured into strengths, growth opportunities, and actionable strategies, supporting teacher reflection and ongoing professional growth.}
  \Description{}
  \label{fig:teaser}
\end{teaserfigure}


\maketitle



\section{Introduction}

Instructional coaching through classroom observation, where trained coaches or peers watch lessons and provide targeted feedback, has long been regarded as one of the most effective strategies for teachers’ professional growth~\citep{ou2018,brown2006}. However, traditional observation-based coaching faces persistent challenges: it is resource-intensive and difficult to scale~\citep{ho2013reliability,kraft2018effect}, often limited to a handful of visits per year; it is shaped by the subjective lens of individual observers, who may vary in their expertise and consistency~\citep{archer2015seeing}; and it captures only fragments of the complex classroom environment, leaving many instructional moves unrecorded or unexamined~\citep{cohen2022change}. The scarcity of qualified coaches further exacerbates these limitations: according to national surveys, a large share of schools either have no instructional coaches or only one or two serving the entire faculty \cite{nces2023, edmb2024}, leaving many teachers without timely or individualized guidance \cite{vermunt2014teacher, unesco2023}.

To address these limitations, researchers have increasingly turned to computational approaches to augment teacher feedback. Recent advances in data analytics and machine learning have enabled systems that generate automated insights into classroom practice or augment teachers reflection~\citep{Jensen2020CHI,alzoubi2021teachactive,demszky2023m,ngoon2024classinsight}. For instance, systems have used machine learning to score instructional quality from transcripts~\citep{Jensen2020CHI}, natural language processing to extract teacher questions~\citep{alic2022computationally, demszky2025automated}, and computer vision to capture classroom activity patterns such as sitting and standing~\citep{alzoubi2021teachactive}. While these tools provide valuable windows into classroom processes, they often remain limited to narrow analytic dimensions and lack the breadth or interpretive richness that human coaching can offer.

Recent advances in modern AI, particularly the emergence of large pretrained foundation models and generative AI technologies~\citep{wei2022emergent}, have created new opportunities for instructional feedback. These models are capable of understanding and reasoning across multimodal contexts (e.g., speech, video, text)~\citep{tsai2019multimodal,comanici2025gemini} and generating outputs that align with human instructions~\citep{ouyang2022training}. Such capabilities enable the vision of an AI instructional feedback system, as illustrated in Figure~\ref{fig:teaser}, that can deliver analytical insights at scale: observing full-length lessons rather than isolated moments, providing comprehensive reviews across multiple pedagogical dimensions, and democratizing access to expert standards and knowledge that are otherwise unevenly distributed~\citep{jurenka2024,ouyang2022training,durante2024agent}.

In this work, we present \textbf{ClassMind}, an AI-enabled instructional feedback system designed to integrate seamlessly into teachers' professional learning routines. Our research is guided by three questions: RQ1: What design features make AI systems effective for supporting classroom observation and teacher professional development? RQ2: How can AI systems be built to deliver reliable and useful feedback on teaching practices? RQ3: How do teachers perceive the usefulness of such systems, and how might they integrate them into their professional workflows?

To investigate these questions, we conducted three types of formative studies: (1) a review of literature and professional reports on classroom observation and feedback, (2) a survey of 75 PK–12 teachers about their practices and expectations for AI-enabled feedback, and (3) semi-structured interviews with six teachers and instructional coaches. From these studies, we distilled five design goals—anchoring feedback in professional standards, ensuring supportive tone, delivering actionable recommendations, and linking feedback to lesson timestamps—that directly reflect teachers’ needs and concerns. Guided by these goals, we developed \textbf{ClassMind}, a system that ingests classroom recordings and generates multimodal annotations and rubric-aligned feedback through AVA-Align, a novel agent framework for temporally precise classroom analysis that outperforms the state-of-the-art general-purpose video agent method on our class feedback generation task.

\textbf{Findings.} Teachers in our user study described \textbf{ClassMind} as useful, easy to use, and novel in its ability to link evidence to frameworks and generate actionable suggestions. They highlighted benefits such as time savings, objectivity, and expanded reflection capacity, while also voicing concerns around cognitive load, privacy, and appropriate role boundaries between AI and human coaches. Importantly, teachers envisioned differential uses across career stages: novices emphasized efficiency and concrete strategies, while veterans sought deeper critique and professional challenge. These findings suggest that AI feedback systems must adapt to users’ expertise, context, and developmental needs.

\textbf{Contributions.} This paper contributes to HCI and learning sciences by:

(1) Presenting the design and implementation of ClassMind, the first open-source platform that uses multimodal generative AI to analyze full-length classroom videos and generate rubric-aligned instructional feedback.

(2) Introducing AVA-Align, a novel technical framework that combines multimodal inputs with pedagogical rubrics and ensures temporal precision, outperforming state-of-the-art video agents on classroom feedback tasks.

(3) Conducting formative studies that ground system design in teachers’ real practices, expectations, and concerns about AI-supported classroom observation.

(4) Offering empirical insights into how teachers perceive AI feedback across different career stages, and discussing challenges such as privacy, cognitive load, and human–AI role boundaries.

Together, these contributions advance understanding of how multimodal generative AI can scale expert coaching while respecting the human-centered values that underpin teacher professional development.

\section{Related Works}




\subsection{Traditional Classroom Observation and Instructional Coaching}

Classroom observation is widely used to both evaluate and improve teaching \cite{hill2013learning}. Systems such as CLASS and the Danielson Framework emphasize classroom management, instructional quality, and student engagement \cite{martinez2016classroom}. While evidence shows that teachers can learn more from observation feedback than from test scores \cite{whitehurst2014evaluating}, reliability remains a challenge, leading many systems to adopt multiple visits and assessors \cite{cantrellKane2013met}.

Instructional coaching—sustained, one-on-one feedback from trained peers or coaches—has emerged as one of the most effective forms of professional development \cite{desimone2017instructional, reddy2021randomized}. Yet its adoption is limited: only 30\% of U.S. schools employ a full-time coach \cite{nces2021coaching}, reflecting the resource-intensive and highly individualized nature of coaching \cite{knight2020cost}. Because coaching is difficult to standardize and scale, its benefits often remain inaccessible to most teachers \cite{blazar2021instructional}.

Together, these limitations highlight a persistent gap: while human observation and coaching deliver valuable, contextualized feedback, they are costly, inconsistent, and hard to scale. Our work addresses this gap by leveraging multimodal AI to generate scalable, rubric-aligned, and context-rich feedback that complements rather than replaces human coaching.

\subsection{Multimodal Foundational AI}

Foundation models are large-scale neural networks pretrained on massive datasets and adaptable to a wide range of downstream tasks with little or no fine-tuning~\cite{Bommasani2021FoundationModels}. Recent advances in large language models (LLMs), such as GPT-4o~\cite{hurst2024gpt}, GPT-o1~\cite{jaech2024openai}, and Claude-4~\cite{anthropic_claude4_systemcard_2025}, demonstrate near–human-level performance in language understanding, instruction following, and long-context reasoning. Building on these advances, researchers have extended LLMs to multimodal inputs, enabling integrated reasoning over audio, vision, video, and other modalities~\cite{han2024onellm,du2025crab,chen2025livecc,li2025lion}. Such models make it possible to interpret rich classroom data streams, including teacher talk, student interactions, and visual activities, through a unified architecture.

Despite this progress, several technical challenges remain highly relevant to classroom observation. First, multimodal models are prone to hallucination, generating inaccurate or ungrounded content~\cite{bai2024hallucination,openai2025hallucination}. Second, instruction following degrades in very long contexts, such as full-length classroom videos~\cite{wu2024longgenbenchbenchmarkinglongformgeneration,wu2024longvideobench}. Third, temporal understanding of video remains limited, making it difficult to pinpoint when specific instructional events occur~\cite{cai2024temporalbench,li2025vidhalluc}.

In parallel, advances in speech technologies have dramatically improved the accuracy of Automatic Speech Recognition (ASR) and speaker diarization. Models such as Whisper~\cite{radford2022whisper,bain2023whisperx} produce accurate transcriptions with word-level timestamps, while diarization systems such as pyannote~\cite{bredin2020pyannote} attribute speech to individual speakers. These capabilities are particularly important in classroom contexts, where distinguishing teacher and student voices under noisy conditions is essential for analyzing discourse.

Our system builds directly on these advances. We employ state-of-the-art multimodal LLMs as the core agent for analyzing classroom video, addressing long-context reasoning, temporal grounding, and multimodal instruction following through our AVA-Align pipeline.

\subsection{AI Systems for Supporting Teacher Growth}
\subsubsection{AI applications on instructional analytics and feedback.} 
AI systems are increasingly used to analyze classroom interactions and generate feedback, offering scalable alternatives to traditional observation and coaching \cite{wang2024artificial}. Recent work has shown that streamlined pipelines integrating teacher self-recording, ASR transcription, and automated discourse feature modeling can achieve moderate accuracy and remain robust to transcription noise \cite{Jensen2020CHI}. In a randomized trial on a large mentorship platform, an NLP feedback system improved instructors’ uptake of student contributions and reduced instructor talk time, with positive effects on learners’ experience and outlook \cite{demszky2023mpowering}. Beyond teacher talk, learning-analytics models detect “uplifting” peer behaviors during small-group work and surface authentic exemplars for reuse in norms-setting \cite{Chandler2024LAK}. 

While these studies focus on classroom discourse and interaction, automated feedback systems more broadly have been categorized into rule-based, data-driven, and hybrid approaches, reflecting a progression from rigid rubrics to adaptive methods \cite{deeva2020review}. These methods extend to open-ended student work such as essays, explanations, and dialogue, enabling scalable formative assessment \cite{huang2023automated,suresh2024fairness,nguyen2025automated}. In large-scale online settings such as MOOCs and programming courses, AI feedback has been shown to alter learning behavior and support persistence, though its impact depends on task structure and learner background \cite{gutl2022edm,yin2022adaptive}.

Instructional coaching also benefits from multimodal evidence. One intervention used automated recognition of posture, gaze, and related signals to augment feedback, yielding durable gains at follow-up \cite{Pereira2021CHI}. Complementary studies leverage multimodal sensing of gaze, turn-taking, and vocal tone to provide real-time prompts that refine social and teaching skills \cite{tsai2020behavioral,guevarra2025llm}. Teachers report finding value in automated indicators such as wait time, uptake of student ideas, and teacher–student talk balance, which surface interaction patterns otherwise invisible during instruction \cite{yun2025customizing}. Real-time systems that flag behavioral anomalies to a human coach further illustrate how AI can complement post hoc analytics \cite{Arakawa2019REsCUE}. However, these systems rely on modality-specific pipelines rather than multimodal large language models (MLLMs) that jointly reason over video, audio, and text.

\subsubsection{AI applications on simulation-based practice for interpersonal skills.} When growth requires rehearsal, large language model role-play offers low-risk spaces to practice mentoring, conferencing, and classroom management with tailored, theory-informed feedback. A perspective articulates an AI Partner or AI Mentor framework that merges experiential practice with formative guidance, making social skill training more accessible and scalable in education and beyond \cite{Yang2024SocialSkillLLMs}. Experimental evidence shows that an LLM conflict simulator grounded in Interest–Rights–Power theory reduces competitive strategies and increases cooperative strategies compared with theory-only training, suggesting that practice plus feedback can shift interpersonal behavior that matters for educator–learner and educator–family interactions \cite{Shaikh2024Rehearsal}.

\section{Formative Study}



To ground our design in teachers’ actual practices and professional development needs, we conducted three formative studies: (1) a narrative review of education literature on classroom observation and instructional feedback, (2) a survey of PK–12 teachers to capture current practices at scale, and (3) semi-structured interviews with teachers and instructional coaches to provide in-depth qualitative perspectives.

\subsection{Narrative Review}

To broaden our perspective beyond HCI and learning sciences, we conducted a narrative review of education-focused media, government reports, publications from international organizations, and NGOs. This review was guided by the central question: \textit{What characterizes high-quality classroom observational feedback?}



Three major conclusions resulted from this exercise:

\emph{(1) Feedback-standard alignment.} Research has demonstrated that observational feedback is most useful when it is anchored in clear and consensual standards of practice. Standards-based frameworks, such as Danielson or CLASS, characterize teaching into more observable indicators. Large-scale implementation studies show that aligning feedback to such standards improves rater agreement and teachers' sense of fairness and usefulness \cite{Garet2017, Levitan2022, oecd2013teachers}, as well as credibility \cite{Cherasaro2016, Frasier2022}. Evaluation experiments that embedded feedback within standards-based observational systems reported gains in instructional practice (and sometimes, student learning) \cite{Steinberg2015, Taylor2012}. 

\emph{(2) Actionable specificity matters.} Teachers consistently prefer feedback that is specific, timely, and immediately useful \cite{Guskey2022, Frasier2022}. For example, repeated, targeted, and multiple feedback cycles can improve instruction \cite{Allen2011, Garet2017, Taylor2012}; providing training that asks observers to be more specific and concrete \cite{Burns2023, Marshall2025}; increasing the length of feedback sessions \cite{Levitan2022}. These practices in turn can elevate teacher self-efficacy \cite{Smith2020, Hattie2009}.

\emph{(3) Feedback deliverance is paramount.} Teachers are most receptive to feedback from observers that are nonthreatening, respectful, collaborative, and demonstrate contextual and subject knowledge \cite{Kerbelyte2018, Frasier2022}. Moreover, while feedback improves practices, trust and perceived observer intent are cornerstones of that success \cite{Will2018, OECD2014}. Schools with strong and healthy professional climates see steeper growth for teachers, as they are submerged in a psychologically safe environment with lifelong learning norms at work \cite{Kraft2014, Randall2020}. This is true even when the general atmosphere is high-pressured; feedback that is constructive and relationship-oriented is more likely to improve teaching \cite{Dee2015}.

\subsection{Online Survey}
To understand current teachers’ use of AI and classroom observation, and their perspectives and potential concerns regarding AI-supported classroom observation, we designed an online survey study.

\textbf{Participants.} We recruited 75 PK–12 teachers via Prolific in the United States (39 female, 36 male; average teaching experience = 12.2 years, median = 10). To capture perspectives across grade levels, half of the participants taught PK–5, with the remainder evenly split across grades 6–8, 9–12, and higher education. Each participant received \$4 compensation. The study was conducted under Institutional Review Board (IRB) approval.

\textbf{Procedure.} After providing informed consent, participants completed an online survey hosted on Prolific. The survey first asked about demographics and teaching background, then explored teachers’ current use of AI tools, followed by their experiences with classroom observation and feedback. Finally, participants reflected on potential opportunities and concerns related to AI-supported instructional feedback.

\textbf{Findings.} 

\emph{AI tool usage}: an overwhelming (92\%) of teachers report using AI of any kind; most frequent is ChatGPT (77\%), followed by Google Classroom (36\%) and Microsoft Copilot (28\%). Most frequently used tasks include lesson planning and creating/adapting materials, while grading and professional development are less common; 

\emph{Classroom observation}: 90\% of teachers report they are formally observed once per year to twice per semester; also, 90\% of observations, whether formal or informal, were done by principals/administration, department chairs, and instructional coaches. Overall, respondents are quite satisfied with all aspects of observations, but limited time for feedback and discussion remains the biggest challenge (45\%), followed by infrequent visits (28\%) and unspecific/useless feedback (27\%);

\emph{Issues with AI}: teachers are most concerned with the ethics of AI (44\%) and accuracy/hallucination in AI (44\%), followed by lack of proper training (32\%). In addition, although 92\% of teachers have used AI at some capacity, only 43\% used it to support their professional growth and development;

\emph{AI feedback willingness}: teachers were most comfortable for AI to feedback their lesson plans (72\% comfortable and above) and suggesting teaching strategies (68\% comfortable and above). Teachers were most hesitant on AI in analyzing student-teacher interactions (52\% uncomfortable or below). While teachers would appreciate how AI can save time (72\%) and provide immediate feedback (59\%), teachers' most major concern lies in privacy/data protection (68\%), lack of human nuance (67\%), and accuracy of AI (61\%). 83\% of teachers would like AI to feedback in the form of a written report, and its most sought-after perspectives of analysis are: differentiation strategies (44\%). content delivery (43\%), and time management (41\%);

\emph{Additional comments}: the open-response section highlighted privacy and security as a recurring concern. Teachers also reflected on their own teaching experience, identity, or curiosity in AI instructional coaching.

\subsection{Expert Interview}

To gain a deeper understanding of how instructional coaching is conducted and to gather rich insights into the potential of AI-supported classroom feedback, we conducted semi-structured interviews with instructional coaches and teachers. The study also explored educators’ beliefs about the possibilities of an AI-based instructional coach.

\textbf{Participants.} We recruited six high school educators via Prolific under IRB approval and informed consent: four teachers and two instructional coaches. Coaches were required to hold a valid teaching license and be actively coaching at Grades 9–12, while teachers were required to hold a valid license, be currently teaching Grades 9–12, and have received instructional coaching within the past year. On average, participants had 17.2 years of teaching experience (median = 17.5, range = 8–25 years). We treat these educators as domain experts given their licensure, extensive professional experience, and active engagement in instructional feedback practices. We focused on Grades 9–12 because COPPA regulations prohibit students under 13 from using generative AI, and many states restrict its use to secondary education. Demographic details are provided in Table\ref{tab:formative_demographics}. Participants received \$60 compensation.

\textbf{Procedure.} Semi-structured interviews were conducted remotely via Zoom in April 2025, each lasting 60–80 minutes. Sessions began with background questions about participants’ roles, schools, and districts, followed by open-ended prompts on how AI could be integrated into teaching and coaching. Interview prompts were guided by the Technology Acceptance Model, emphasizing perceived ease of use and perceived usefulness, adapted into qualitative rather than Likert-scale items.

\textbf{Analysis.} Interviews were transcribed and thematically coded [20] using NVivo 15. Through open coding, we identified both cross-cutting themes across participants and themes emerging uniquely within individual narratives.

\textbf{Findings.} Our thematic analysis revealed five major themes concerning teachers’ and coaches’ perspectives on their current instructional feedback practices and their attitudes toward AI-enabled instructional feedback.

\textit{Attitudes and trust:} participants described mixed experiences with human coaching, with several teachers reporting “normal or bad” relationships that reduced the perceived value of feedback. Against this backdrop, they responded positively to the idea of AI assistance, with two coaches saying they “loved it” and saw “a lot of things it could apply to,” though they stressed that trust would depend on careful framing and endorsement: “The question will never be what AI can do, but how to present AI in a way that is palatable.”

\textit{Coverage and subject expertise:} teachers noted that human coaches often observe only fragments of lessons—sometimes “just five minutes”—and may lack subject-specific knowledge, limiting their ability to provide high-quality feedback; they saw AI as potentially overcoming these limitations by capturing entire class sessions and drawing on broader knowledge resources.

\textit{Practicality and efficiency:} Participants valued AI’s potential to streamline repetitive tasks such as compiling hundreds of yearly observation forms and to provide timely lesson-level insights, with one teacher remarking, “Yes, give me another point of view to teach it.” Others stressed the importance of immediacy: “I need something that I can implement right away in the next class.” At the same time, teachers cautioned that too much feedback could become overwhelming: “Everything you do can be nit-picked … no time to sit with the knowledge if I’m being bombarded all the time.” Together, these perspectives highlight the dual imperative for AI systems to accelerate efficiency while carefully managing cognitive load.

\textit{Human versus AI roles:} coaches emphasized that “90\% of coaching is done outside formal sessions” in trust-building interactions that AI cannot replicate, but consistently agreed that AI excels at analytic tasks such as quantifying student participation, surfacing classroom patterns, and tying observations to professional frameworks.

\textit{Feedback priorities and formats:} all participants identified student engagement as the “absolute primary focus,” followed by delivery quality and higher-order thinking, and expressed preferences for feedback that was concrete, observable, and efficiently presented (e.g., bullet points, charts, or timestamped video clips). Some also desired customizable features such as toggling between evaluation models or viewing exemplar videos of others’ practice, though the latter raised confidentiality challenges. In sum, participants envisioned AI as a supportive tool that could capture fuller classroom contexts, reduce administrative work, provide subject-relevant insights, and deliver timely, concrete feedback, while leaving relational aspects of coaching to human coaches.

\begin{table*}[h]
\centering
\begin{tabularx}{\textwidth}{
    >{\arraybackslash}X
    >{\arraybackslash}X
    >{\arraybackslash}X
    >{\arraybackslash}X
    >{\arraybackslash}X
    >{\arraybackslash}X
    >{\centering\arraybackslash}X
    >{\centering\arraybackslash}X
}
\toprule
\textbf{ID} & \textbf{Gender} & \textbf{Ethnicity} & \textbf{State} & \textbf{Subject} & \textbf{School Type \& Title I} & \textbf{Years Teaching} & \textbf{Years Coaching} \\
\midrule
IC1 & Male   & White    & KY & STEM           & Independent, Yes   & 20 & 15 \\
IC2 & Male   & White    & CT & Math           & Technical, Yes     & 25 & 17 \\
T1  & Female & White    & WI & English        & Comprehensive, Yes & 15 & N/A \\
T2  & Female & White    & PA & Social Studies & Comprehensive, Yes & 23 & N/A \\
T3  & Male   & White    & FL & Social Studies & Charter, No        & 8  & N/A \\
T4  & Female & Hispanic & CA & Biology        & Independent, Yes   & 12 & N/A \\
\bottomrule
\end{tabularx}
\caption{Demographic and professional background of participants.}
\label{tab:formative_demographics}
\end{table*}

\section{Design Goals}

Drawing on insights from our literature review, survey, and semi-structured interviews with teachers and instructional coaches, we scoped a prototype designed to provide static feedback on single high school lessons. From this formative phase, we derived five design goals that guided system development, each directly addressing recurrent challenges identified in the data. Although features such as interactive coaching and longitudinal progress tracking emerged as promising directions, we deferred them to future work due to scope constraints and open technical challenges, as discussed further in Section~\ref{Sec:limitations}.

\textbf{DG1: Anchor Feedback in Established Teaching Practices and Professional Standards.}  
Feedback should be explicitly tied to recognized frameworks (e.g., Danielson, CLASS) and observable instructional practices to enhance credibility and fairness. Evidence from our narrative review shows that framework-linked feedback improves fairness, credibility, and instructional outcomes \cite{Garet2017, Levitan2022, Cherasaro2016, Steinberg2015}. Coaches we interviewed similarly emphasized that they would trust the AI system more when it “ties observations to professional frameworks,” and survey data revealed that some of teachers’ top-requested support areas—such as teaching strategies, differentiation strategies, and questioning techniques—are all codified in professional standards, underscoring the need for clear framework linkage.

\textbf{DG2: Communicate Feedback in a Supportive and Encouraging Tone.}  
The system should prioritize constructive, nonthreatening, and strengths-oriented language to foster trust and receptivity. Prior research highlights that teachers respond best to nonthreatening, respectful, and collaborative feedback, particularly when delivered in psychologically safe environments \cite{Kerbelyte2018, Will2018, Kraft2014}. Expert interviews reinforced this, with participants contrasting “bad” coaching relationships with the need for constructive feedback and stressing that trust depends on “how to present AI in a way that is palatable.” 

\textbf{DG3: Deliver Specific and Immediately Actionable Recommendations.}
Feedback should move beyond description to provide concrete next steps that teachers can directly incorporate into their practice. Prior research consistently highlights the importance of specificity and timeliness for effective feedback \cite{Guskey2022, Allen2011, Burns2023}. Our survey results similarly revealed that most teachers expect AI to save time and deliver immediate insights. In interviews, several teachers emphasized their preference for feedback that offers alternative teaching strategies or suggestions they could implement in their very next class.

\textbf{DG4: Link Feedback to Lesson Timestamps for Efficient Navigation.}  
Standards-based frameworks emphasize observable, evidence-linked practices, which naturally lend themselves to timestamped anchoring. Survey results showed that while most teachers preferred written reports (83\%), a large majority also valued time-saving features (72\%), underscoring the potential of timestamp-linked navigation to improve efficiency. Expert interviews reinforced this point: teachers frequently expressed frustration that coaches often observed “just five minutes” of a lesson or held “vague conversations without evidence.” In contrast, timestamped, full-lesson coverage was viewed as a way for AI to provide richer, evidence-grounded insights while enabling teachers to efficiently revisit specific moments of their instruction.

\textbf{DG5: Support Consistent Analysis of Full-Length Classroom Videos.}  
Research shows that multiple, repeated, and extended feedback cycles improve instructional practice\cite{Taylor2012, Garet2017}, which in turn requires observation of full lessons rather than fragments. Survey respondents reported frustration with limited feedback time (45\%) and infrequent visits (28\%), and teachers in interviews noted that human coaches often miss most of a lesson. Accordingly, AI systems should be capable of capturing entire sessions and providing analysis with a holistic view.

\section{System Implementation}

Guided by the design goals, we implemented ClassMind, an AI-powered system that analyzes classroom recordings and generates rubric-aligned feedback for teachers. The system is designed to help teachers reflect on their practice by combining objective annotations of classroom activity with evaluative comments grounded in professional standards.

The implementation proceeds in three layers. First, we adopt established educational taxonomies (e.g., Danielson, Bloom, COPUS) as a conceptual blueprint to ensure that generated feedback is interpretable and pedagogically meaningful. Second, we introduce AVA-Align, our agent framework for long-video understanding, which aligns multimodal evidence with rubric criteria. Finally, we describe the product functionality and user flow, showing how teachers interact with the prototype to upload recordings, navigate annotations, and receive actionable feedback.

\subsection{Established Taxonomies}

To guide the feedback generation workflow, we grounded our system in established educational taxonomies that are widely recognized in classroom observation research. This choice was motivated by insights from our expert interviews, where teachers and instructional coaches emphasized the importance of aligning feedback with well-known frameworks. These frameworks provide shared vocabularies and structured criteria that ensure AI-generated feedback is interpretable, credible, and pedagogically meaningful.

\textbf{Danielson’s Framework for Teaching.} The Danielson Framework articulates a standards-aligned language for evaluating K–12 instruction across four domains: Domain 1 - Planning and Preparation, Domain 2 - Classroom Environment, Domain 3 - Instruction, and Domain 4 - Professional Responsibilities, further divided into 22 sub-domains~\cite{danielson2013fft,danielsongroup_fft_web}. In our system, Danielson serves as the primary rubric for aligning feedback, ensuring that AI-generated comments resonate with teachers’ professional standards and evaluative contexts.

\textbf{Bloom’s Taxonomy of Educational Objectives.} Bloom’s revised taxonomy classifies cognitive objectives into a hierarchy: (1) Remember, (2) Understand, (3) Apply, (4) Analyze, (5) Evaluate, and (6) Create~\cite{bloom1956taxonomy,anderson2001taxonomy}. We use Bloom’s taxonomy to analyze the distribution of teacher questions in classroom videos, distinguishing higher- from lower-order questioning. This enables the system to highlight patterns in cognitive demand and provide teachers with actionable insights on discourse practices.

\textbf{Classroom Observation Protocol for Undergraduate STEM (COPUS).} COPUS codes classroom interactions into observable categories such as lecturing, questioning, and group work~\cite{smith2013copus,cwsei_copus_web}. Within our system, COPUS informs the objective activity distribution visualizations, allowing teachers to see how instructional time is allocated and to contextualize feedback within patterns of classroom engagement.

\subsection{Supporting AI Functionalities}
Supporting AI functionalities are summarized in table \ref{tab:supporting-functionalities}.

\begin{table*}[h]
\centering
\caption{Overview of Supporting AI Functionalities Implemented in the System}
\begin{tabular}{p{3cm}p{5cm}p{6cm}}
\toprule
\textbf{Functionality} & \textbf{Implementation} & \textbf{Notes / Limitations} \\
\midrule
Transcription \& Captioning 
& Whisper-Large-v3 + Pyannote 3.1 diarization; Gemini-2.5-Flash for visual captions (2-min segments). 
& $\sim$30 sec per 30-min video on A6000; Teacher vs. students only (no individual student IDs); captions not shown to users but used downstream. \\
\addlinespace
Activity Annotation 
& Video-Language Model aligned with COPUS codes (teacher lecturing, writing, Q\&A, student listening, etc.). 
& Granularity: sentence-level; multiple co-occurring activities allowed; higher temporal resolution than 2-min fixed interval coding. \\
\addlinespace
Question Distribution 
& Two-step pipeline: (1) LLM extracts questions from transcript, (2) model classifies via Bloom’s taxonomy (Table~\ref{tab:bloom}). 
& Justification provided per category; wait-time analysis possible but left as future work. \\
\addlinespace
Outline Generation 
& Merge transcripts and captions into structured summary. 
& Provides quick recap; generally aligns with human summarization practices. \\
\addlinespace
Video Recommendation 
& Retrieval pipeline: (1) LLM generates search query, (2) E5-Large retrieves top-10, (3) LLM re-ranks/filter. 
& Private db (2100+ TeachingChannel videos); not redistributed; standard coarse-to-fine retrieval setup. \\
\bottomrule
\end{tabular}
\label{tab:supporting-functionalities}
\end{table*}

\subsection{AVA-Align Agent Framework for Video Understanding}

In the context of classroom observation, our technical task is: given a long classroom video (typically 20–60 minutes) and a rubric, generate feedback items that are (1) grounded in video evidence, (2) temporally accurate, and (3) aligned with pedagogical standards. Achieving this goal requires addressing three key challenges:

\begin{enumerate}
    \item Long-context video understanding. Classroom recordings often span 10–120 minutes, exceeding the context limits of most video–language models. Even state-of-the-art systems struggle with hallucinations and information loss on long sequences~\citep{shu2025video}.
    \item Temporal grounding. Meaningful feedback requires second-level accuracy about when events occurred. Off-the-shelf video–language models remain unreliable in producing precise timestamps~\citep{li2025mitigating,wang2025videoitg}.
    \item Instruction following with multimodal context. When both video input and textual rubrics are combined, models frequently fail to maintain alignment and adhere to rubric-based instructions~\citep{li2024multi,ratzlaff2025training}.
\end{enumerate}

\begin{figure*}[h]
  \centering
  \includegraphics[width=\linewidth]{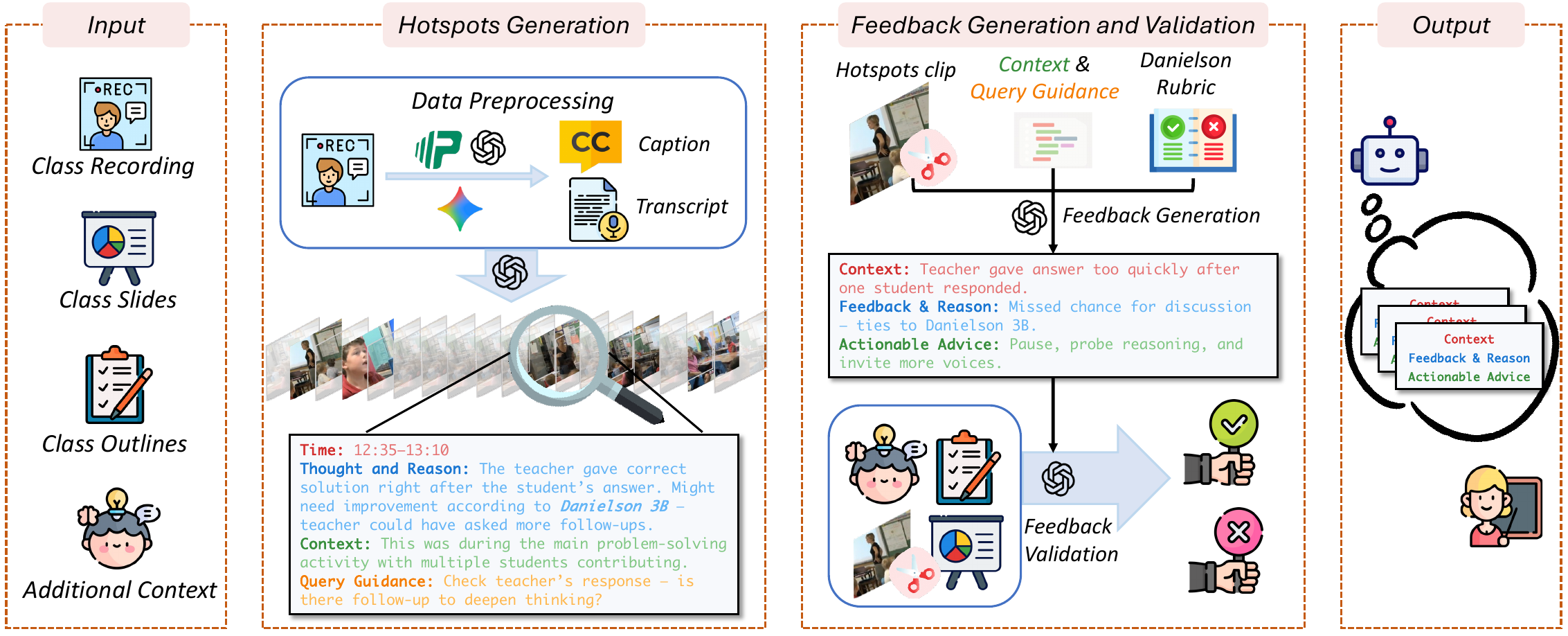}
  \caption{\textbf{Illustration of Our AVA-Align Video Agent Framework.} Classroom inputs include recordings (preprocessed into transcripts and captions), slides, outlines, and context, from which the system identifies “hotspots.” Each hotspot is analyzed with rubric guidance to generate context-grounded feedback, which is then validated against the video. The output provides teachers with rubric-aligned comments that include context, reasoning, and actionable advice.}
  \Description{}
  \label{fig:agent-workflow}
\end{figure*}

To address these issues, we introduce AVA-Align (Adaptive Video Agent with Alignment for long rubrics), our core agent architecture for generating rubric-aligned feedback from classroom recordings. AVA-Align is designed to ensure temporal precision and faithful rubric alignment. Figure~\ref{fig:agent-workflow} illustrates the workflow.

\textbf{Input Processing.} We begin by dividing videos into 2-minute segments, a window size that balances efficiency and accuracy for the Gemini-2.5-Flash video foundation model. Each segment is captioned and synchronized with transcripts, producing second-level, timestamped descriptions of the entire class session. This serves as the foundation for downstream analysis.

\textbf{Hotspot Generation.} To focus model attention, we operationalize the Danielson Framework~\citep{danielson2013fft} as a structured rubric. Each rubric dimension includes threshold criteria and example indicators at different evaluation levels. Prior work shows that large language models are inconsistent in generating free-form feedback~\citep{shrivastava2024measuring}; grounding the generation process in rubrics mitigates this issue and aligns outputs with professional standards~\citep{hashemi2024llm}.

Captions and transcripts are then merged in temporal order, and the model identifies potential hotspots—segments where strengths or weaknesses may emerge in relation to rubric dimensions. Each hotspot is accompanied by contextual information to ensure that subsequent reasoning accounts for the broader lesson, not just isolated clips.

\textbf{Feedback Generation and Validation.} For each hotspot, general text descriptions are insufficient to capture subtleties such as student engagement or teacher gestures. To reduce cognitive load, we narrow the focus to one video segment and one rubric dimension at a time. The language model first generates targeted guidelines (e.g., “check whether students respond actively to the teacher’s prompts”), which direct the video–language model’s attention.

Next, the model produces rubric-aligned feedback, integrating segment-level evidence with relevant rubric criteria. Each feedback item is refined to improve granularity, ensure rubric consistency, and emphasize actionable insights.

Finally, a validation step leverages the video–language model to cross-check each feedback item against its timestamped video evidence. Building on findings that models are more reliable at validation than generation~\citep{libenchmarking}, this consistency check strengthens factuality and fidelity to classroom events.

\subsection{System Outputs and Interaction}
\label{sec:product-flow}

The user flow begins when a teacher uploads a classroom recording. Teachers may optionally provide contextual materials (e.g., lesson plans or notes) to enrich interpretation. Once processing begins, the system generates two categories of outputs: objective annotations that describe classroom activity, and rubric-aligned feedback that provides evaluative insights.

\begin{figure*}
    \centering
    \includegraphics[width=1\linewidth]{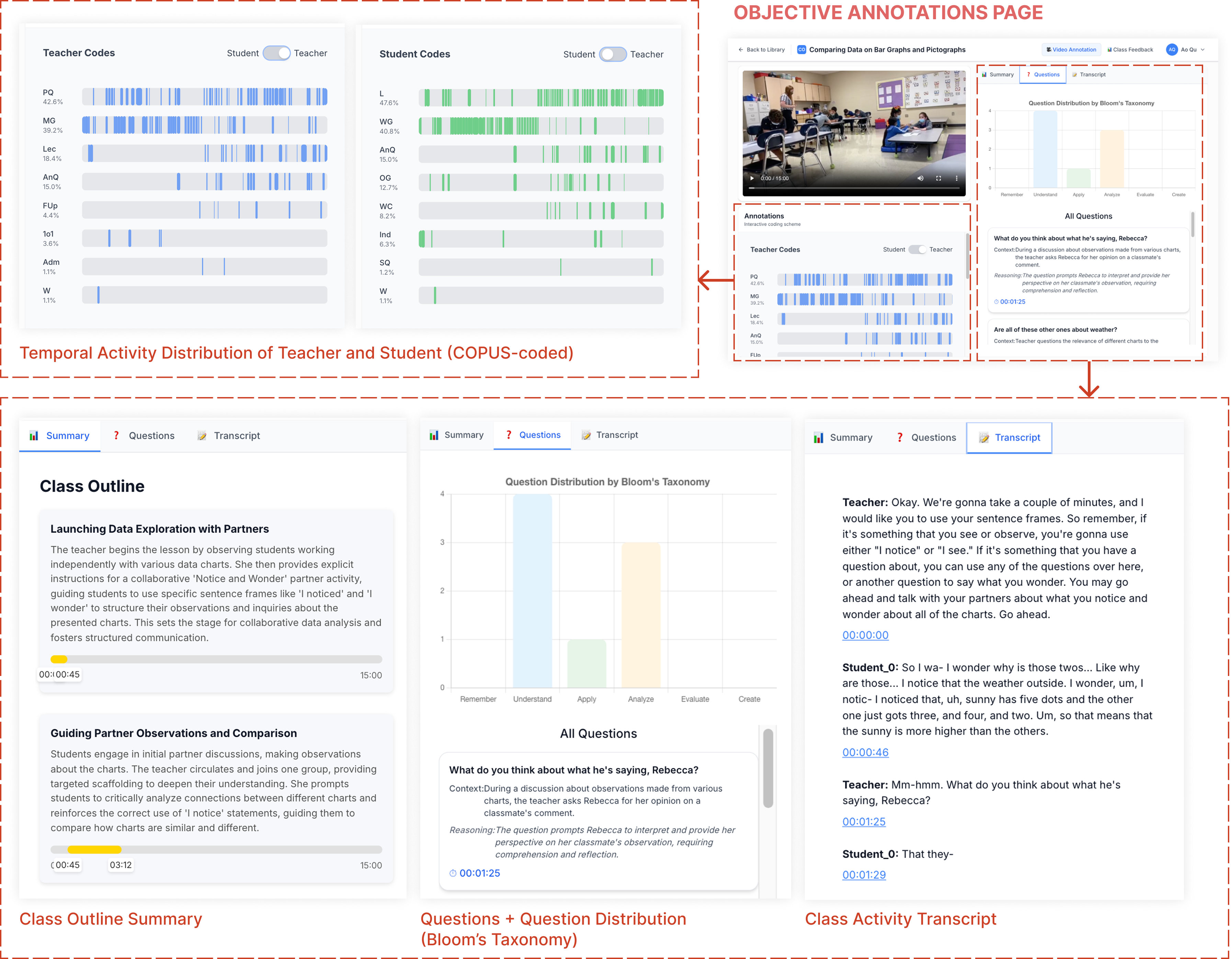}
    \caption{\textbf{Objective annotations generated by ClassMind.} The system provides (a) temporal activity distribution of teacher and student interactions (COPUS-coded), (b) question distribution categorized by Bloom’s taxonomy, (c) class outline summaries, and (d) transcripts labeling teacher and student turns. These outputs help teachers reconstruct lesson events, navigate recordings efficiently, and contextualize feedback.}
    \label{fig:system1}
\end{figure*}

\textbf{Objective annotations. }These outputs help teachers reconstruct what occurred during their lesson. The system visualizes activity distribution using COPUS codes, showing the proportion of time spent lecturing, questioning, or group work. Hovering over a label reveals its description, and clicking navigates directly to the relevant video segment. The system also produces a question distribution categorized under Bloom’s taxonomy, with each question linked to its timestamp and accompanied by rationale for the classification. Finally, transcripts and a lesson outline summarize the overall session, labeling speakers as teacher or student and providing a high-level structure of the class. Together, these annotations enable efficient navigation and contextual review. (See Figure \ref{fig:system1}.)

\begin{figure*}
    \centering
    \includegraphics[width=1\linewidth]{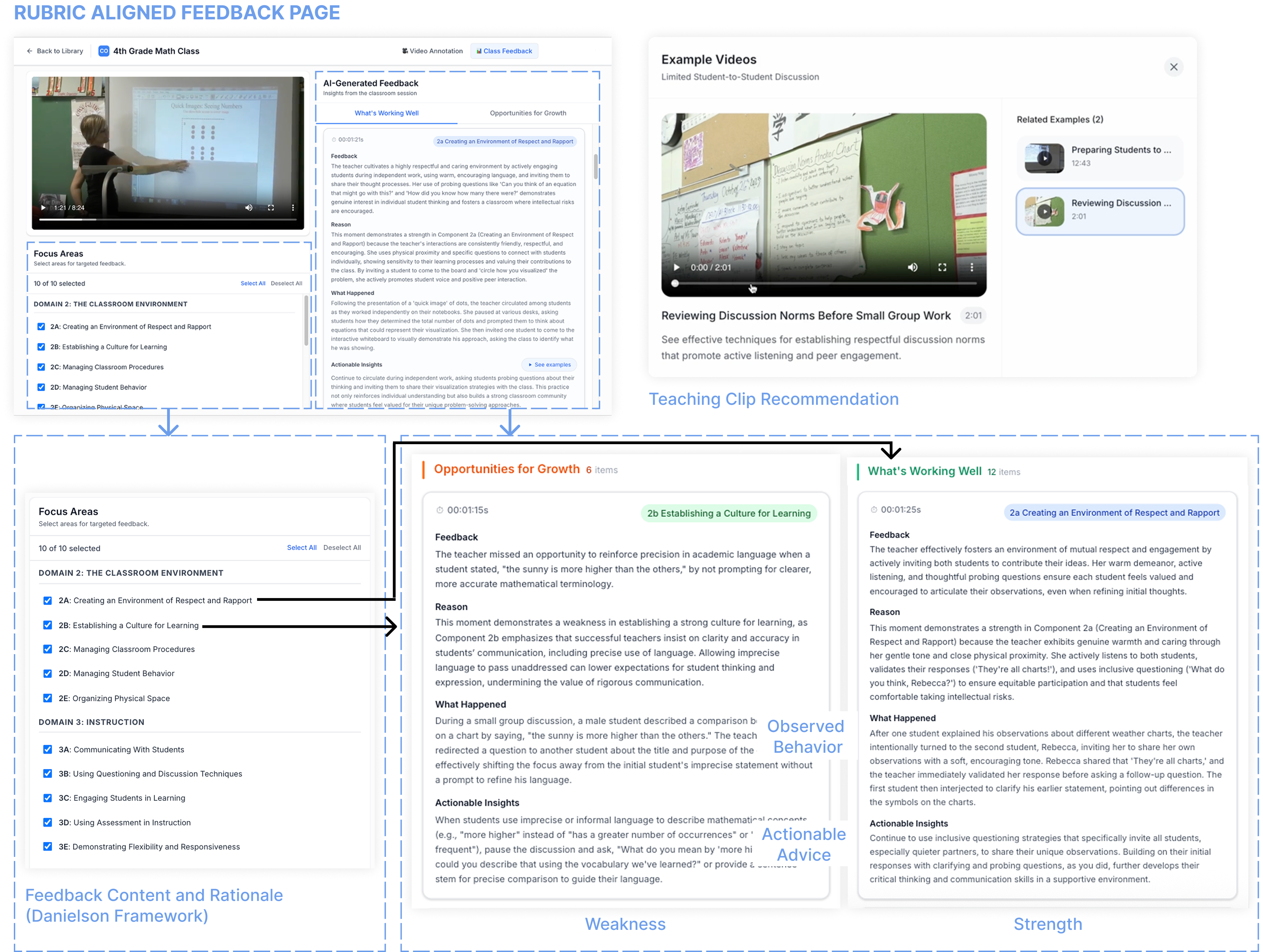}
    \caption{\textbf{Rubric-aligned feedback generated by ClassMind.} Building on the AVA-Align pipeline, the system produces evaluative comments aligned with pedagogical rubrics. In our prototype, we use the Danielson Framework, though alternative rubrics are supported. Feedback is structured into three components: (a) feedback content and rationale grounded in rubric dimensions, (b) observed behaviors with evidence from the class video, and (c) actionable advice suggesting alternative strategies. To contextualize recommendations, the system also (d) retrieves example teaching clips that illustrate relevant practices.
}
    \label{fig:system2}
\end{figure*}

\textbf{Rubric-aligned feedback.} Building on the AVA-Align pipeline, the system generates feedback items aligned with pedagogical rubrics. In our prototype we adopt the Danielson Framework, though the system also supports alternative or customized rubrics. Each feedback item is structured to first highlight strengths, then surface weaknesses, and finally offer actionable advice for improvement. Entries include three components: (1) feedback content and rationale, (2) observed behaviors of the teacher during class activity, and (3) actionable advice for future practice. To further ground recommendations, the system retrieves example video clips from the class that illustrate relevant teaching moments, accompanied by short explanations of their pedagogical significance. (See Figure \ref{fig:system2}.)

\section{Technical Evaluation\label{Sec:eval_ai}}

We evaluated the core AI functionalities of our system across four tasks: video feedback generation, teacher question detection and classification, activity annotation, and speaker diarization. Standard transcription and summarization modules were not extensively evaluated, as they adopt best-practice implementations with well-established benchmarks.

\subsection{Video Feedback}

We benchmarked our video agent pipeline against two representative baselines, Gemini-2.5-Pro and Deep Video Discovery (see Appendix~\ref{app:baselines}). The evaluation focused on reliability, usefulness (significance and feasibility), and temporal coverage (definitions in Appendix~\ref{app:metrics}). We conducted experiments on 10 videos from the \textbf{ATLAS video library} and reported results with standard deviations where appropriate in Table~\ref{tab:numeric_results}. Our system demonstrated higher factuality, more pedagogically meaningful insights, and a more balanced distribution of feedback across lessons.

\begin{table*}[h]
\centering
\caption{Video feedback evaluation results}
\label{tab:numeric_results}
\begin{tabular}{lcccc}
\toprule
\textbf{Model} & \textbf{Factuality (\%) $\uparrow$} & \textbf{Significance (1--5) $\uparrow$} & \textbf{Feasibility (1--5) $\uparrow$} & \textbf{Temporal Coverage (Entropy) $\uparrow$} \\
\midrule
\textbf{Ours (Video Agent)}   & \textbf{100} & \textbf{4.40 $\pm$ 0.66} & \textbf{4.50 $\pm$ 0.50} & \textbf{0.85 $\pm$ 0.07} \\
Gemini-2.5-Pro~\citep{comanici2025gemini}       & 66.7 & 3.00 $\pm$ 0.76 & 3.11 $\pm$ 0.45 & 0.23 $\pm$ 0.05 \\
Deep Video Discovery~\citep{zhang2025deep} & 94.3 & 3.61 $\pm$ 0.71 & 3.89 $\pm$ 0.74 & 0.84 $\pm$ 0.06 \\
\bottomrule
\end{tabular}
\end{table*}

\subsection{Other Core Modules}
Beyond video feedback, we evaluated three additional modules:  (1) Teacher Question Detection and Classification: Our system achieved an F1 score of 0.964, a recall of 0.93, and a precision of 1.  
(2) Activity Annotation: Grounded in COPUS but adapted for sentence-level segmentation, the system achieved a Micro-F1 of 0.89 for teacher activities, and Micro-F1 of 0.78 for student activities.  
(3) Speaker Diarization: Our diarization module achieved a Jaccard Error Rate (JER) of 10.3\%. Definitions of all metrics are provided in Appendix~\ref{app:metrics}.

Overall, our system demonstrated strong reliability, balanced temporal coverage, and pedagogically meaningful feedback. It also showed robust performance across question detection, activity annotation, and diarization, supporting its feasibility for scalable instructional feedback.

\section{User Studies: Simulated-User Role-Play with Classroom Observation Video}

Our study aimed to investigate how teachers experienced an AI-driven instructional feedback prototype \textbf{ClassMind} within a controlled, role-play scenario. Specifically, we asked \textbf{RQ3: How do teachers perceive the usefulness of such systems, and how might they integrate them into their professional workflows?} This guided our study design and analysis, foregrounding teachers’ lived experiences rather than system performance metrics.

\subsection{Participants}

We recruited six secondary-school teachers in the United States via Prolific, drawn from a prior survey where they had indicated interest in follow-up studies. Eligibility required current employment in an official school in the US. Participants varied in teaching experience (1 - 20 years), subject areas, and school contexts across different tiers of the U.S. education system, with both gender representation (Female = 4, Male = 2). This diversity enabled us to capture a range of perspectives on instructional reflection and AI adoption. The study was approved by the \textbf{Institutional Review Board (IRB)}. All participants provided written informed consent and received USD \$60 compensation for their time.

\begin{table*}[h]
\centering
\caption{Participant Demographics.}
\label{tab:participants}
\begin{tabular}{lccccc}
\hline
\textbf{ID} & \textbf{Gender} & \textbf{Ethnicity} & \textbf{Subject} & \textbf{Years Teaching}  & \textbf{Video Selection} \\
\hline
P1 & Male & White & Spanish & 20 & Facilitating Explicit Phonics \\
P2 & Male & White  & Maths & 5 & Comparing Data on Bar Graphs and Pictographs \\
P3 & Female & White  & English Literature & 9 & The Veldt by Bradbury\\
P4 & Female & White  & Maths & 20 & Comparing Data on Bar Graphs and Pictographs\\
P5 & Female & Asian  & Science & 2 & Addressing issues of genetic manipulation\\
P6 & Female & Asian  & ELA & 1 & Facilitating Explicit Phonics\\
\hline
\end{tabular}
\end{table*}

\subsection{Study Materials and Context}

Due to ethical and logistical constraints in obtaining classroom recordings involving minors, we drew on the \textbf{ATLAS video library} \citep{NBPTS_ATLAS_2025} (\textit{Accomplished Teaching, Learning, and Schools}). ATLAS is an online, password-protected repository maintained by the National Board for Professional Teaching Standards (NBPTS), featuring authentic, unedited classroom video cases of National Board Certified Teachers. Each case includes a classroom video, the teacher’s written reflection, instructional materials, and extensive indexing to teaching and learning frameworks. It covers grades PreK-12, multiple subject areas, and diverse school settings. ATLAS was a practical choice because all U.S.-registered teachers can apply for access, and its videos (10--15 minutes) are concise enough to allow a holistic understanding of classroom dynamics without overwhelming participants.

Each participant selected a classroom video from ATLAS in a subject domain most familiar to their own teaching practice. This ensured that the material was relevant to their professional context, while also maintaining consistency in format and accessibility.

In addition to video footage, ATLAS provides instructional commentary, including details on student dynamics, lesson planning rationale, teacher actions, and after-class reflection. This supplementary information mirrors the kind of contextual data teachers might normally supply to an AI system alongside classroom recordings. Incorporating these materials allowed us to simulate a more authentic use case, where the AI has access to both video evidence and rich instructional context.

\subsection{Study Procedure}

The study followed a \textbf{simulated-user role-play design}, in which teachers engaged with the AI coaching prototype as if it were analyzing their own classroom practice. The procedure consisted of four phases:

\textbf{Role Induction.} One day before the study session, participants watched a 10--15 minute ATLAS video they had selected in their own subject domain. They were prompted to imagine themselves as the teacher in the classroom and to submit a short reflective log responding to: \textit{``What areas of improvement or feedback would be most valuable to you in teaching this lesson?''} This step established a baseline reflection without AI support and ensured role immersion.

\textbf{System Exploration.} At the start of the session, participants interacted with the AI coaching prototype for 10 minutes of unguided exploration, using a ample classroom video distinct from their chosen main video. This phase familiarized participants with the interface and elicited initial usability impressions.

\textbf{Personalized Feedback Evaluation.} Participants then reviewed the AI-generated feedback on the video they had selected earlier. They were asked to treat the feedback as though it applied directly to their own teaching practice. This phase was self-paced, with participants indicating when they felt their review was complete.

\textbf{Post-Study Measures.} After the task, participants completed standardized questionnaires (USE, NASA-TLX, FATE) and took part in a 20--30 minute semi-structured interview. The interview probed perceptions of the AI’s support for reflection, trustworthiness, fairness, emotional responses, and potential integration into teaching routines. Interviews were recorded and transcribed for later analysis.

\begin{table*}[t]
\centering
\caption{Study Phases Overview.}
\label{tab:timeline}
\begin{tabular}{llp{7cm}}
\toprule
\textbf{Phase} & \textbf{When} & \textbf{Activities} \\
\midrule
Role Induction & Day --1 (self-paced) &
Watch self-selected ATLAS video + commentary; submit baseline reflection log \\
System Exploration & T0 (10 min) &
Unguided exploration with sample video (not main); initial impressions (USE) \\
Personalized Feedback Evaluation & T1 (self-paced) &
Review AI feedback on selected ATLAS video as if it were own class \\
Post-Study Measures & T2 (20--30 min) &
Complete USE, NASA-TLX, FATE; semi-structured interview (recorded) \\
\bottomrule
\end{tabular}
\end{table*}

\subsection{Measures}

We employed a combination of standardized self-report instruments and qualitative data sources to capture participants’ experiences with \textbf{ClassMind} .

\subsubsection{Standardized Questionnaires}

Usability was assessed with the \textbf{Usefulness, Satisfaction, and Ease of use  Questionnaire (USE)} ~\citep{Lund2001USE}, which measures perceived usefulness, satisfaction, ease of use, and ease of learning. Cognitive workload was measured using the \textbf{NASA Task Load Index (NASA-TLX)} ~\citep{grier2015tlx}. Perceptions of responsible AI qualities were evaluated with the \textbf{Fairness, Accountability, Transparency, and Explainability Questionnaire (FATE)} ~\citep{shin2021Explainability}, which includes constructs such as causability, fairness, accountability, transparency, explainability, personalization, trust, satisfaction, emotion, and convenience.

\subsubsection{Qualitative Data}

Prior to the study session, participants submitted \textbf{baseline reflection logs} based on their chosen ATLAS classroom video. These logs provided insight into how teachers normally reflect on classroom practice in the absence of AI support. Following the main task, participants completed a semi-structured interview (20--30 minutes). Interviews probed teachers’ perceptions of the AI’s support for reflection, alignment with their teaching practices, trust and fairness concerns, emotional experiences, and anticipated use in professional routines. All interviews were audio-recorded and transcribed verbatim for subsequent analysis.

We intentionally relied on self-report instruments and qualitative accounts to foreground participants’ lived experiences with the system. This methodological choice reflects our interest in teachers’ interpretive judgments, rather than system telemetry or performance metrics.

\section{Results}

We present findings in relation to the RQ3 outlined in Introduction. Data were drawn from standardized questionnaires (USE, NASA-TLX, FATE) and semi-structured interviews, with qualitative themes complementing descriptive statistics. Given the small sample ($N = 6$), results are \textit{exploratory} and intended to surface experiential patterns rather than population-level effects.

\subsection{Perceptions of Usability, Workload, and Responsible AI Qualities}

\subsubsection{Usability (USE)}

Participants rated the system as \textbf{highly usable overall} ($M = 5.9/7$, 95\% CI [5.5, 6.4], $\alpha = .89$; Table~\ref{tab:USE}). \textit{Ease of Learning} was strongest ($M = 6.7$), followed by \textit{Ease of Use} ($M = 6.1$), while \textit{Usefulness} ($M = 5.6$) and \textit{Satisfaction} ($M = 5.3$) were more tempered. Internal consistency was good across subscales ($\alpha \approx .79$--.91).

Teachers described the system as easy to pick up—one noted she ``\textit{figured it out within minutes}’’—and appreciated direct navigation to salient classroom moments. At the same time, several remarked that some analytic categories were ``\textit{not immediately clear}’’ if unfamiliar with the underlying frameworks. This pattern helps contextualize why \textit{Ease of Learning} approached ceiling, whereas \textit{Usefulness} and \textit{Satisfaction} varied with background and immediate needs. In addition, a novice teacher (P5) explicitly de-prioritized classroom observation relative to current goals (lesson content and planning). Although she rated the system as effective for the feedback task, she indicated limited immediate utility for her priorities. A sensitivity check suggested that this single case attenuated the \textit{Usefulness} and \textit{Satisfaction} means.

\begin{table*}[h]
\centering
\caption{USE questionnaire results ($N = 6$). Subscale means, SDs, and 95\% t-based CIs (1–7 scale; Cronbach’s $\alpha$ per subscale (item counts in parentheses). }
\label{tab:USE}
\begin{tabular}{lccccc}
\toprule
Subscale        & Mean & SD   & 95\% CI (lwr--upr) & \% of Max & Cronbach’s $\alpha$ \\
\midrule
Ease of Learning & 6.7 & 0.2 & [6.3, 7.0] & 96\% & .91 \\
Ease of Use      & 6.1 & 0.3 & [5.7, 6.5] & 87\% & .85 \\
Usefulness       & 5.6 & 0.5 & [5.0, 6.2] & 80\% & .79 \\
Satisfaction     & 5.3 & 0.6 & [4.6, 6.0] & 76\% & .82 \\
\midrule
\textbf{USE Total} & \textbf{5.9} & 0.4 & [5.5, 6.4] & 84\% & .89 \\
\bottomrule
\end{tabular}
\end{table*}

\subsubsection{Workload (NASA-TLX)}

Self-reported workload was \textbf{modest overall} (Raw TLX Overall: $M = 28.5/100$, 95\% CI [18.1, 38.9]; Table~\ref{tab:TLX}). \textit{Mental Demand} ($M = 42$) and \textit{Effort} ($M = 37$) were relatively higher, while \textit{Physical Demand}, \textit{Temporal Demand}, and \textit{Frustration} were consistently low (all $M$s < 20). Participants rated \textit{Performance} favorably ($M = 23$, where higher = poorer).

Teachers described this workload as stemming primarily from cognitive sense-making. As P6 explained, ``\textit{most of the effort was making sense of the AI suggestions and connecting them to my goals}.'' Others emphasized the absence of stress or time pressure, with one remarking that it ``\textit{didn’t feel rushed}.'' These perspectives underscore that the tool itself was not burdensome; rather, the work involved interpreting how to act on its feedback. Thus, workload was \textbf{manageable}, with effort concentrated in reflection rather than in navigation.

\begin{table}[h]
\centering
\caption{Unweighted NASA-TLX results ($N = 6$). Subscale means, SDs, and 95\% \emph{t}-based CIs (0--100 scale)}
\label{tab:TLX}
\begin{tabular}{lccc}
\toprule
Subscale        & Mean & SD   & 95\% CI (lwr--upr) \\
\midrule
Mental Demand   & 42 & 18 & [28, 56] \\
Physical Demand & 12 &  7 & [ 5, 19] \\
Temporal Demand & 18 &  9 & [10, 26] \\
Performance     & 23 & 11 & [14, 32] \\
Effort          & 37 & 15 & [24, 50] \\
Frustration     & 14 &  6 & [ 8, 20] \\
\midrule
\textbf{TLX Overall} & \textbf{28.5} & 12 & [18, 39] \\
\bottomrule
\end{tabular}
\end{table}

\subsubsection{Responsible AI Qualities (FATE)}

Perceptions of responsible AI were \textbf{positive overall} ($FATE_{Total} M = 5.8$, 95\% CI [5.4, 6.2], $\alpha = .88$; Table~\ref{tab:FATE}). Among core constructs, \textit{Fairness} ($M = 6.2$) and \textit{Causability} ($M = 6.1$) were highest, while \textit{Accountability} ($M = 5.3$) and \textit{Explainability} ($M = 5.4$) were somewhat lower but still favorable. Outcome ratings were also strong, particularly \textit{Emotion} ($M = 6.0$) and \textit{Satisfaction} ($M = 5.9$).

Qualitative evidence reinforced these results. One teacher P4 emphasized that the feedback ``\textit{didn’t feel biased against my teaching style},'', ``\textit{no stereotypes},'' aligning with high fairness scores. At the same time, several noted uncertainty about responsibility and interpretability, asking ``\textit{who is responsible if the AI misreads a situation?}'' or requesting ``\textit{more detail on why something was flagged}.'' Such comments suggest that while teachers found the system impartial and generally trustworthy, they also desired more transparency about its evaluative processes. In short, teachers perceived the system as \textbf{fair and reliable}, but looked for stronger \textbf{accountability and explainability} to fully support its use in professional settings.

\begin{table*}[h]
\centering
\caption{FATE results ($N = 6$). Subscale means, SDs, and 95\% \emph{t}-based CIs (1--7 scale); Cronbach’s $\alpha$ per subscale (item counts in parentheses).}

\label{tab:FATE}
\begin{tabular}{lccccc}
\toprule
Subscale        & Mean & SD   & 95\% CI (lwr--upr) & \% of Max & Cronbach’s $\alpha$ \\
\midrule
Fairness         & 6.2 & 0.4 & [5.8, 6.6] & 89\% & .87 \\
Causability      & 6.1 & 0.5 & [5.6, 6.6] & 87\% & .83 \\
Accountability   & 5.3 & 0.6 & [4.6, 6.0] & 76\% & .78 \\
Transparency     & 5.5 & 0.5 & [5.0, 6.0] & 79\% & .80 \\
Explainability   & 5.4 & 0.7 & [4.7, 6.1] & 77\% & .81 \\
\midrule
Performance      & 5.7 & 0.5 & [5.2, 6.2] & 81\% & .82 \\
Convenience      & 5.8 & 0.4 & [5.4, 6.2] & 83\% & .85 \\
Trust            & 5.6 & 0.6 & [5.0, 6.2] & 80\% & .84 \\
Satisfaction     & 5.9 & 0.5 & [5.4, 6.4] & 84\% & .83 \\
Emotion          & 6.0 & 0.4 & [5.6, 6.4] & 86\% & .86 \\
\midrule
\textbf{FATE Core}     & \textbf{5.7} & 0.5 & [5.2, 6.2] & 81\% & .88 \\
\textbf{FATE Outcomes} & \textbf{5.8} & 0.4 & [5.4, 6.2] & 83\% & .87 \\
\textbf{FATE Total}    & \textbf{5.8} & 0.4 & [5.4, 6.2] & 83\% & .88 \\
\bottomrule
\end{tabular}
\end{table*}

\subsection{Integration of AI Feedback into Teacher Reflection and Growth}

We examined how teachers engaged with AI-generated observations and feedback, focusing on reflection practices and perceived professional growth. We triangulated pre-session reflections with reactions to outputs (Bloom's taxonomy distributions, transcripts, Danielson-coded feedback, and timeline navigation). With a small sample $(N=6)$, findings are exploratory and surface how trust, usefulness, and autonomy play out across developmental stages.

\subsubsection{Objectivity as a Baseline for Trust}

Teachers consistently grounded their trust in the system's ability to produce objective evidence about classroom practice. All teachers in our study highlighted Bloom's taxonomy as one of the most compelling elements of the system. While Bloom's is a foundational framework familiar to every teacher, its adoption in this context allows it to act not simply as charts, but as a mirror of gaps between intention and practice. As P6 noted, teachers may ``think'' they are using Bloom's consistently, but this system made visible whether this was actually the case. Similarly, P3 reflected, \textit{``sometimes I do a lot of thinking for the students \dots{} this could give myself feedback on what I am actually asking kids to do.''} Furthermore, trust deepened when outputs aligned with what teachers had noticed. P3 immediately recognized a student repeatedly trying to interject and reacted, \textit{``that's the student! \dots{} I was like wow, this (the system) picked it up.''} Precision also reduced rewatching costs, as P5 used timeline jumps to locate issues without screening an entire video. However, unfamiliar frameworks might reduce interpretability. P3 reported not understanding the Danielson domain labels despite agreeing with the substance, and transcripts degraded during multi-speaker discussion, which marked a boundary condition. Overall, teachers trusted the system most when feedback was grounded in shared professional language and when evidence cohered with their in-class perceptions.

\subsubsection{From Vague Noticing to Articulated Reflection}

Teachers reported that the system helped them move from tacit impressions to explicit, analyzable reflections. For novices, this often meant putting words to an intuition that something felt off. As P5 explained, \textit{``the system helps me to organize and express my thinking.''} By grounding observations in transcripts and categorized questioning patterns, the AI turned vague knowing into structured evidence that could be examined and discussed. Teachers also contrasted this experience with the generic feedback they typically receive from supervisors, who may not have time to sit through an entire lesson. P4, a math teacher, described being evaluated by a district-level science supervisor who oversees many teachers, often observing for only 10--15 minutes per year and completing a checklist with broad judgments such as ``good'' or ``not good.'' Whereas, the system identified a subject-specific imprecise use of a mathematical term, which P4 praised as a meaningful catch and a common pitfall among elementary teachers without a strong math background. This contrast illustrated how the system's detailed feedback could provide meaningful support to teachers. Across participants, the system was experienced as specific and actionable. Examples included concrete rewording suggestions for classroom talk (P3) and indicating quieter students (P4). P6 emphasized that the analysis provided a clear, objective, and rational bridge from theory emphasized in teacher education to practical classroom decisions. Altogether, the system expanded reflective capacity by transforming transient or vague impressions into analyzable accounts that supported reasoning about why particular moments mattered and what to change next.

\subsubsection{Tensions Around Autonomy and Cognitive Load}

Comprehensiveness can be viewed as both strength and burden. Several teachers welcomed breadth but reported cognitive overload. P6 initially felt validated by positive recognition, then discouraged by a long list of improvements. P1 pointed to instances that felt nitpicky, such as a dropped marker being flagged as a classroom management issue. Teachers responded by curating the feedback. As P2 put it, \textit{``you can always use what you need \dots{} and discard what doesn't fit your class.''} Experienced teachers described scanning the holistic capture to locate what mattered and ignoring the rest, while novices asked for compressed summaries, for example one- or two-sentence takeaways with brief rationale. Preferences for minimum coverage also varied, with P4 expecting at least some signal for every domain and P6 preferring less. Effective integration depended on regulating granularity in order to preserve a sense of agency.

\subsubsection{Divergent Needs of Novice and Experienced Teachers}

Career stage moderated interpretation and adoption. Novices (P5, P6) valued actionable tips yet were sensitive to tone and volume. P5 emphasized the importance of feedback phrased in a supportive manner and noted that classroom analysis may be deprioritized when immediate curriculum planning dominates. Experienced teachers (P3, P4) actively sought critique to challenge entrenched habits, saw department-level applications such as supporting new teachers, and compared the AI favorably to time-pressed supervisors, also noting perceived fairness and objectivity around issues like gender and ethnicity. Frequency intentions reflected this split. Some teachers envisioned periodic self-audits, for example every 2 to 3 weeks, while others imagined systemic roles for evaluation evidence, long-term tracking, and targeted coaching. The same feedback profile was reassuring for novices when gentle and brief, and energizing for experienced teachers when critical and comprehensive, which suggests a need for adaptive framing.

\section{Discussion}
The findings from our study illuminate not only the opportunities but also the tensions that arise when AI is introduced into teacher professional development. While teachers saw promise in AI’s ability to extend feedback, save time, and complement existing supports, they also raised critical concerns about ethics, privacy, tone, and role clarity. These perspectives highlight that AI feedback systems cannot be understood in isolation; rather, they must be situated within broader ecosystems of human mentorship, institutional practices, and the realities of teachers’ workload and career stages. In this discussion, we draw together these insights to explore four themes: the evolving partnership between AI and human coaches, the ethical and privacy challenges that shape adoption, the design practices that make AI feedback actionable and trustworthy, and the pragmatic differences in how such tools are perceived across teacher career stages and educational contexts.
\subsection{Human–AI Coaching Partnerships}
Our study naturally brings attention to what the relationship should be between AI and human coaches. Teachers often described AI as excelling in analytics, coverage, and efficiency, offering more concrete insights than the generic feedback they typically received. One participant noted, \textit{``This is way better than what supervisors currently do.''} In contexts where access to mentors is limited, AI was seen as especially valuable—particularly \textit{``for new teachers … who don’t have colleagues,''} or in districts where \textit{``there’s just not enough people to go around.''} From the perspective of administrators and coaches, AI also emerged as a value add: it could take workload off routine observations, provide evidence-based reports, and free human coaches to focus on deeper professional growth. This mirrors how companies such as TeachFX and Edthena build their platforms—they position themselves as analytical support and offer features specifically designed to augment instructional coaches’ capacity~\citep{TeachFX,EdthenaAI_Coach_For_Teachers}.

At the same time, participants emphasized what AI cannot replace. Human coaches contribute mentorship, trust, and professional judgment that extend beyond observation. As one teacher put it, \textit{``AI can’t know every student or personality.''} Teachers also expressed concern that AI may lack adaptability over time: \textit{``It will always have the same standard … humans can adapt and build better understanding.''} This critique resonates with broader concerns in the HCI and education literature that algorithmic systems tend toward rigid standardization~\citep{williamson2020objectivity}, while human feedback evolves with context, institutional culture, and individual teacher growth.

Interestingly, several participants described AI as providing a \textit{``safe layer of objectivity.''} By reducing interpersonal pressures that sometimes come with supervisory authority, AI feedback was often perceived as easier to accept. This echoes research on fairness and trust in algorithmic systems~\citep{lee2018understanding,madaio2020co,afroogh2024trust}. Yet, as~\citep{williamson2020objectivity} reminds us, objectivity itself is socially constructed, raising questions about how teachers negotiate the authority of AI-generated judgments over time.

Taken together, our findings point toward a partnership model~\citep{holstein2019co,liu2024social,cheng2025transitioning}. AI can expand access to feedback, reduce administrative burden, and deliver scalable, evidence-linked observations, while humans remain indispensable for mentorship, adaptability, and relational care. This model aligns with a broader trend in human–AI interaction research, where scholars highlight the division of labor as central to trustworthy collaboration~\citep{shneiderman2020human,yang2020re,shao2024collaborative}. Industry practice shows a similar pattern: while some startups market \textit{``AI coaching replacements,''} the tools that gain traction are those that integrate AI as a partner, extending analytic capacity while leaving space for the relational dimensions of professional growth~\cite{holmes2019artificial,TeachFX,betterup2025}. Taken together, these observations invite reflection on how future systems might embody partnership in ways that are both practical and sustainable.

\subsection{Ethical and Privacy Concerns}

Across formative insights and user studies, our data converge on a unified concern that teachers have about ethics and privacy in AI for education. In the survey, privacy and data protection were the top-cited concerns around AI-enabled feedback. Formative interviews elaborated that trust hinges on framing, endorsement, and role clarity for AI versus humans, with participants emphasizing risks of over-collection and overload alongside potential benefits. User study interviewees underscore a stance of vigilant pragmatism: \textit{``take everything with a grain of salt,''} \textit{``always employ critical thinking,''} and a generational skepticism toward AI (\textit{``maybe I’m older, I don’t trust AI''}), while even enthusiastic early-career teachers reported blending AI inputs with personal judgment. Together, these findings indicate that privacy expectations, consent, and institutional accountability are prerequisites for adoption.

These findings align with recent theoretical work that frames generative AI in schools as an ethics-and-privacy problem organized around children’s rights, purpose limitation, informed consent, role-based access, and transparent data flows. Systematic reviews argue that responsible adoption requires governance beyond technical fixes, including policy guardrails and curricular attention to ethics, privacy literacy, bias mitigation, and age-appropriate explainability \cite{Gouseti2025EthicsK12, GarciaLopez2025EthicalRegulatory, Ma2025ResponsibleAILiteracy, WenTian2025AIK12Policy}. Design guidance for child–AI co-creation similarly prioritizes privacy by default, developmentally appropriate disclosures, and constraints that reduce hallucinations and over-reliance, translating abstract principles into product-level requirements for tools used by minors \cite{Cai2025ChildAICoCreation}.

Practice-oriented studies complement this picture. Teachers report under-preparedness and hesitation, often confining generative AI to back-office tasks that minimize exposure of student data, and call for explicit standards and policy support to scaffold ethical classroom use \cite{Cheah2025GenAIK12, Chiu2024GenAIPractices, Ng2025GenAIStrategies}. Research with middle schoolers reveals curiosity coupled with misconceptions about tools like ChatGPT, underscoring the need for consent processes and disclosures that are developmentally tuned to student understanding \cite{Belghith2024MiddleSchoolers}. Work on undisclosed use shows that transparency is not a formality but a lived ethical issue that shapes trust and participation, echoing our participants’ insistence on clarity and choice \cite{Adnin2025Disclosure}. Studies of academic integrity find no immediate surge in cheating post ChatGPT, shifting emphasis from policing to values-aligned policy and instructional redesign, which teachers in our interviews also endorsed as the credible path forward \cite{Lee2024CheatingHS}. Taken together, the triangulated evidence and literature specify actionable requirements for AI-supported observation and feedback in K–12, that is, \emph{narrowly scoped purposes, least-data pipelines with on-device or ephemeral processing where feasible, role-based access with educator control, age-appropriate disclosure and opt-in consent, population-sensitive bias auditing}, and \emph{ongoing ethics-and-privacy learning} for both teachers and students.

\subsection{Reflections on Practices for AI Feedback}
A recurring theme in our findings is that good AI feedback is not only about accuracy, but also about trust, tone, and actionability. Prior work shows that trust in automation depends not only on correctness but also on appropriate scope and framing~\citep{lee2004trust,yang2020re}. Teachers in our study similarly stressed that AI comments should be grounded in observable evidence: \textit{``most of them are based on the video \dots there’s nothing the AI create by themselves.''} This reinforces prior findings that explainability and verifiability are central to trust~\citep{shneiderman2020human}. Tone was equally consequential. Encouraging, strengths-first phrasing fostered receptivity, whereas deficit-focused or overly harsh tone risked alienation: \textit{``If a brand new teacher read this, I think they would get a little upset \dots you do want the good feedback, but in a nice way.''} The affective part of feedback, even when generated by AI, aligns with educational research on the role of feedback valence in learning~\citep{hattie2007power,demszky2023m} and HCI work on affective framing~\citep{grawemeyer2017affective}. Yet tone alone was insufficient. Teachers repeatedly emphasized actionability: \textit{``That’s what teachers want. They want to know what they’re doing well and what they can do to improve.''} This resonates with feedback research underscoring that actionable guidance predicts uptake~\citep{narciss2013designing,van2015effects}.  

In the generative AI era, where 56\% of our survey participants reported using AI for creativity and idea generation as one of the primary use cases, this points to a broader principle. Detailed observation is valuable only when coupled with clear next steps or scaffolding, and teachers even imagined AI going further by suggesting new activities.  

Finally, the tension between granularity and conciseness shaped perceptions of usefulness. Some praised AI’s exhaustive detail: \textit{``the class feedback is very detailed \dots the principal or the mentor don’t have time to do this.''} Others found long lists overwhelming: \textit{``Oh my God, that’s so long. I don’t wanna see it.''} This suggests design solutions such as progressive disclosure (for example, summaries first with expandable details) and bundling insights into actionable recommendations~\citep{hearst2009search,springer2018progressive}. Taken together, these findings underscore that effective AI feedback must balance trustworthiness, encouraging tone, actionable guidance, and digestible presentation in order to translate into meaningful changes in practice.

\subsection{Pragmatic Insights Across Career Stages and Education Scenarios}
A key theme in our findings is that \textit{who} the feedback is intended for---and \textit{why}---profoundly shapes what kind of feedback is useful. Novice and early-career teachers, often overwhelmed by workload, expressed a strong preference for efficiency supports and concrete, actionable guidance. Veteran teachers, by contrast, sought deeper insight, critique, and reflection to refine established practices. As one teacher noted after 20 years of experience: \textit{``I like \ldots actionable insights into questioning and discussion \ldots I would definitely act upon that.''} This progression suggests that AI systems should adapt both the tone and depth of feedback to teachers' experience levels. This observation aligns with prior HCI and education research: co-design studies of AI curricula highlight that teachers' values, workflows, and levels of expertise fundamentally shape adoption. For instance, \citet{lin2021engaging} found that novice teachers emphasized logistical and direct supports, whereas more experienced teachers sought tools with greater conceptual or ethical depth. Similarly, \citet{tan2025artificial} shows that professional development initiatives around AI often fail to differentiate supports by experience level.  

Beyond career stage, participants also envisioned broader scenarios of use. Some saw value in teacher education programs or even for students themselves: \textit{``I could even see it being valuable for students to do \ldots like, you know, you record yourself giving a presentation and then it gives you feedback.''} A novice teacher similarly reflected: \textit{``I can use it when I was in school taking teacher training.''} These insights resonate with a broader HCI literature on AI-mediated social skills coaching, such as conversational agents for public speaking or autism support~\citep{jiang2023simulation,tang2024emoeden}, and with commercial applications like Yoodli, which provides automated presentation feedback~\citep{yoodli}. While real classroom deployments may raise privacy concerns, such tools in teacher training contexts could offer direct and immediate value.  

Finally, teachers consistently drew a clear boundary between professional development and evaluation. They welcomed AI as a reflective partner but resisted its use as an administrative judgment tool. This distinction mirrors findings in prior research: when technologies are framed as instruments of managerial surveillance, they can trigger concerns about privacy and provoke adversarial dynamics~\citep{terpstra2023online,kwapisz2024privacy,ahn2021co,roemmich2023emotion}.

\section{Limitations and Future Work\label{Sec:limitations}}

Our study surfaces several important limitations that also point toward promising future directions. (1)\textbf{Study Context and Deployment}. In our user study, teachers did not watch their own classroom videos, which meant that some features, such as user-specified context setting and video recommendation clips, were not fully tested. Teachers lacked contextual information beyond their own class to meaningfully specify settings, and participants could not view recommended clips because the system was built on a commercial database without access rights. To address these limitations, we are pursuing partnerships with schools to enable deployed studies in authentic settings, allowing districts to curate recommendation databases internally and supporting longitudinal investigations that track teacher development over time. (2)\textbf{Toward Interactive Coaching}. Beyond technical deployment, teachers also expressed a desire for more open-ended, dialogic interactions rather than one-way feedback. Realizing this vision requires AI systems that are more proactive and conversationally adaptive, capable of supporting reflective dialogue and collaborative planning without being overly submissive. As one of the early works applying multimodal generative AI in this domain, we chose to first establish a foundation of technological maturity, leaving interactive coaching capabilities, such as conversational planning, collaborative goal-setting, and iterative feedback exchange, to future work. (3)\textbf{Instruction and Content Expertise}. Finally, while our current focus has been on instructional practices (e.g., questioning techniques, pacing, student engagement), teachers also highlighted the importance of content-specific feedback. For example, they described situations where instructional coaches lacked sufficient subject-matter expertise to give actionable advice on how to explain technical concepts. AI holds potential to fill this gap by integrating subject-level knowledge with instructional guidance, offering more holistic support. Future research should explore how to combine content-sensitive and instructional analysis in ways that complement the strengths of human coaches.

\bibliographystyle{ACM-Reference-Format}
\bibliography{references}

\clearpage

\appendix

\section{Baselines and Metrics}
\label{app:baselines}

\subsection{Baseline Models}
\begin{itemize}
    \item \textbf{Gemini-2.5-Flash}: A state-of-the-art multimodal foundation model optimized for video captioning and temporal reasoning.
    \item \textbf{GPT-4o}: A large multimodal model widely used for general-purpose video--language understanding.
    \item \textbf{Deep Video Discovery}: A research pipeline that applies long-context video embedding and retrieval to identify salient events.
\end{itemize}

\subsection{Evaluation Metrics and Experiment Details\label{app:metrics} }
We introduce the definitions of the evaluation metrics below. Our agent was evaluated against two baseline approaches on ten teaching videos from the ATLAS video library \citep{NBPTS_ATLAS_2025}, ranging from 8 to 25 minutes in length. In addition, one teacher and one instructional coach rated the outputs on significance and feasibility, without being informed of the underlying method. For class activity annotation, we only labeled three videos due to its labor-intensive nature.

\begin{itemize}
    \item \textbf{Factuality}: Proportion of feedback items that accurately reflected classroom events.
    \item \textbf{Significance}: Extent to which feedback targeted pedagogically meaningful aspects of instruction and offered actionable insights.
    \item \textbf{Feasibility}: Degree to which feedback could be integrated into teachers’ workflows without undue burden.
    \item \textbf{Temporal Coverage}: To evaluate how evenly feedback was distributed across a lesson video, we computed normalized temporal entropy of the feedback timestamps. First, the full video duration was divided into equal time bins, and the duration or count of feedback events in each bin was aggregated to form a probability distribution. Shannon entropy $H=-\sum_i p_i \ln p_i$ was then calculated and normalized by the maximum possible entropy $\ln(k)$, where $k$ is the number of bins. The resulting normalized entropy $H_{\text{norm}}=H/\ln(k)$ ranges from 0 (all feedback concentrated in a single moment) to 1 (feedback perfectly balanced across the lesson), providing a robust measure of temporal coverage.``` 
    \item \textbf{F1}, \textbf{Recall}, \textbf{Precision}: Standard classification metrics used for teacher question detection.
    \item\textbf{Micro-F1}: A standard measure for multi-class annotation performance, computed by aggregating true positives, false positives, and false negatives across all classes before calculating precision and recall. This emphasizes overall accuracy and reflects the system’s performance on all activity instances.
    \item \textbf{Jaccard Error Rate (JER)}: $JER = 1 - \frac{|A \cap B|}{|A \cup B|}$, where $A$ and $B$ are sets of speaker segments from system output and ground truth. Lower values indicate stronger agreement.
\end{itemize}

\begin{figure*}[t]
  \centering
  \begin{subfigure}[t]{0.95\linewidth}
    \centering
    \includegraphics[width=\linewidth]{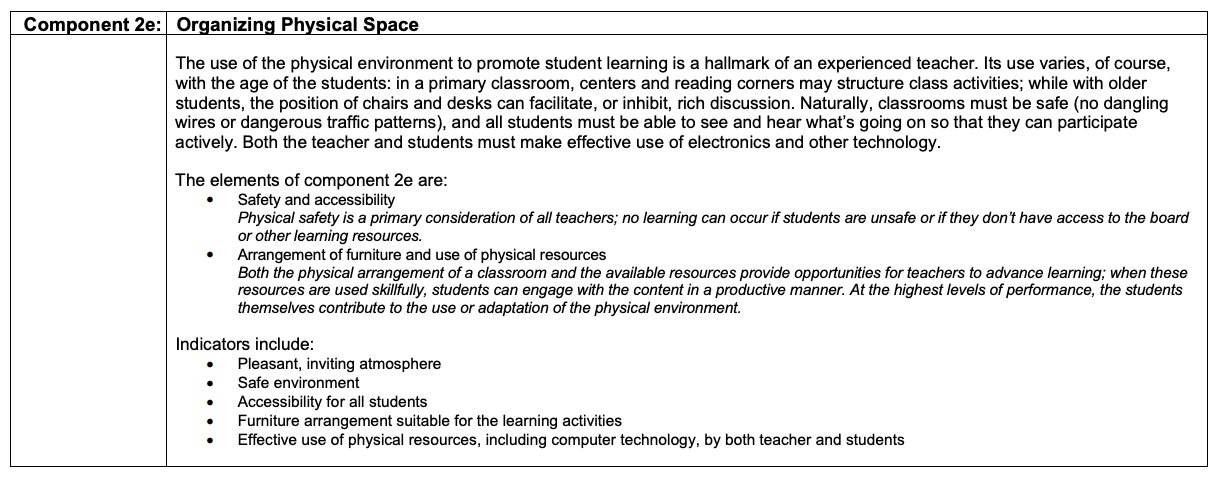}
    \caption{Performance levels for Component 2e (Organizing Physical Space), showing Unsatisfactory, Basic, Proficient, and Distinguished descriptors with critical attributes and possible examples.}
    \label{fig:danielson-2e-levels}
  \end{subfigure}
  \vspace{0.5em}

  \begin{subfigure}[t]{0.95\linewidth}
    \centering
    \includegraphics[width=\linewidth]{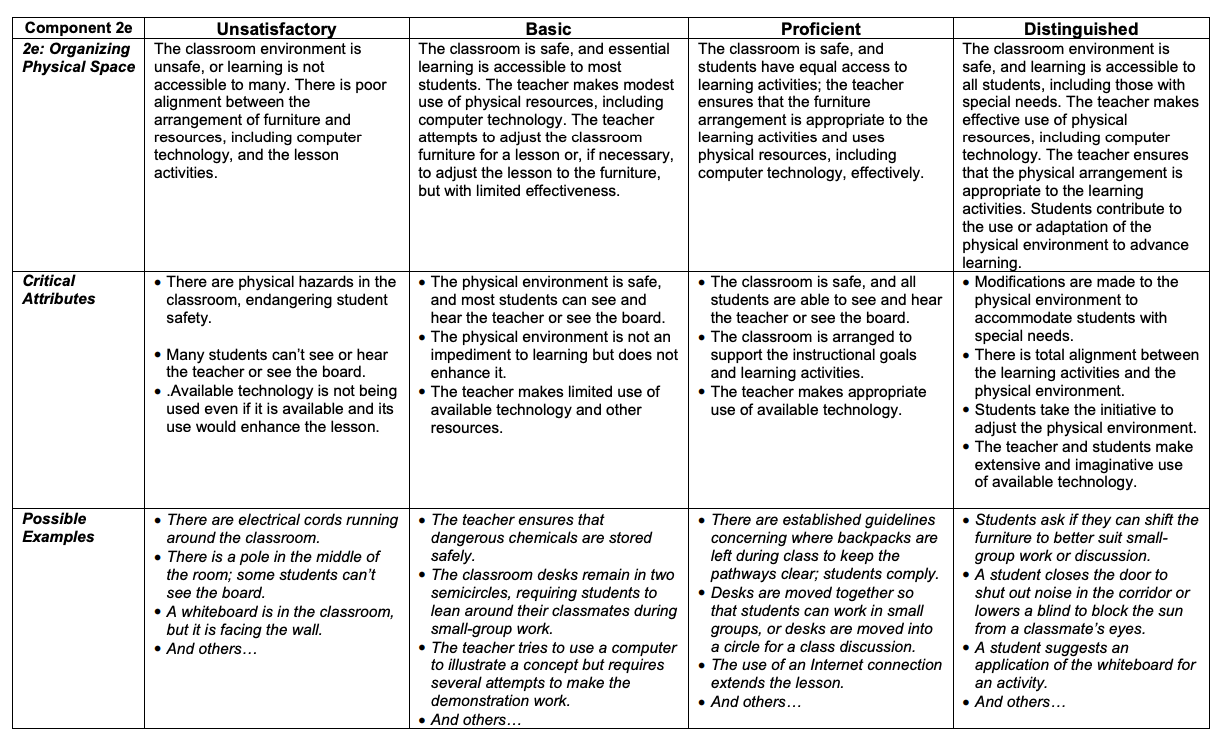}    
    \caption{Overview of Component 2e, including elements (safety and accessibility; arrangement of furniture/resources) and key indicators of effective practice.}
    \label{fig:danielson-2e-elements}
  \end{subfigure}

  \caption{Danielson Framework, Component 2e (Organizing Physical Space). These rubrics define expectations for safe and effective classroom environments, which we used to ground our system’s feedback design.}
  \label{fig:danielson-2e}
\end{figure*}

\begin{table}[t]
\small
\setlength{\tabcolsep}{6pt}
\renewcommand{\arraystretch}{1.25}
\centering
\caption{Bloom’s taxonomy: cognitive levels with example actions.}
\label{tab:bloom}
\begin{tabularx}{\linewidth}{@{}p{0.9cm}p{3cm}Y@{}}
\toprule
\textbf{Level} & \textbf{Cognitive Process} & \textbf{Description and Example Verbs} \\
\midrule
6 & Create & Use existing information to make something new.  
\emph{Invent, develop, design, compose, generate, construct} \\
\addlinespace
5 & Evaluate & Make judgments based on sound analysis.  
\emph{Assess, judge, defend, prioritize, critique, recommend} \\
\addlinespace
4 & Analyze & Explore relationships, causes, and connections.  
\emph{Compare, contrast, categorize, organize, distinguish} \\
\addlinespace
3 & Apply & Use existing knowledge in new contexts.  
\emph{Practice, calculate, implement, operate, use, illustrate} \\
\addlinespace
2 & Understand & Grasp the meaning of something.  
\emph{Explain, paraphrase, report, describe, summarize} \\
\addlinespace
1 & Remember & Retain and recall information.  
\emph{Reiterate, memorize, duplicate, repeat, identify} \\
\bottomrule
\end{tabularx}
\end{table}

\begin{figure}[h]
\centering
\includegraphics[width=0.85\linewidth]{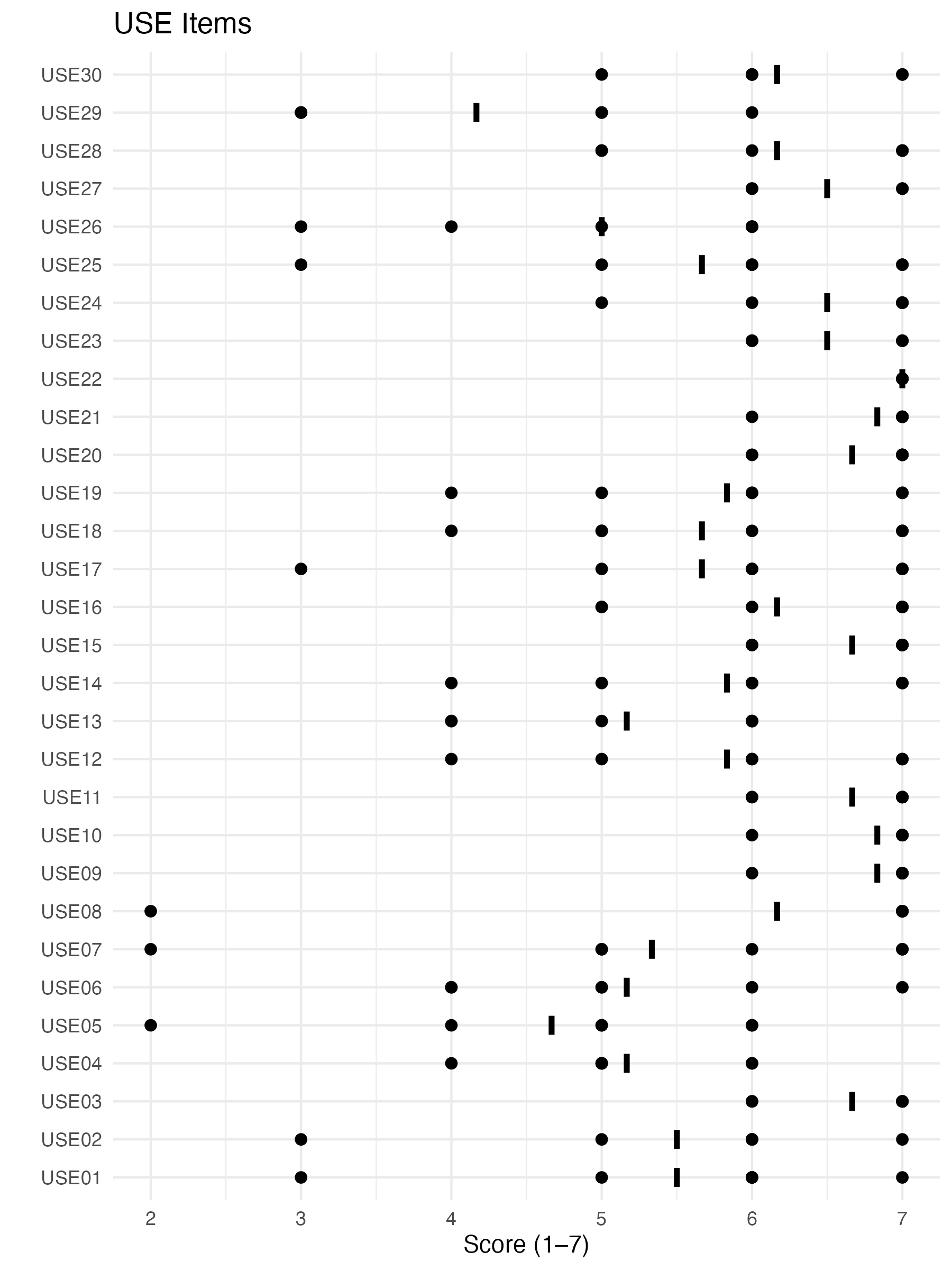}
\caption{USE Questionnaire subscales (per-participant points + mean crossbar). Error bars represent 95\% CIs for descriptive purposes only ($N=6$). }
\label{fig:USE}
\end{figure}

\begin{figure}[h]
\centering
\includegraphics[width=0.85\linewidth]{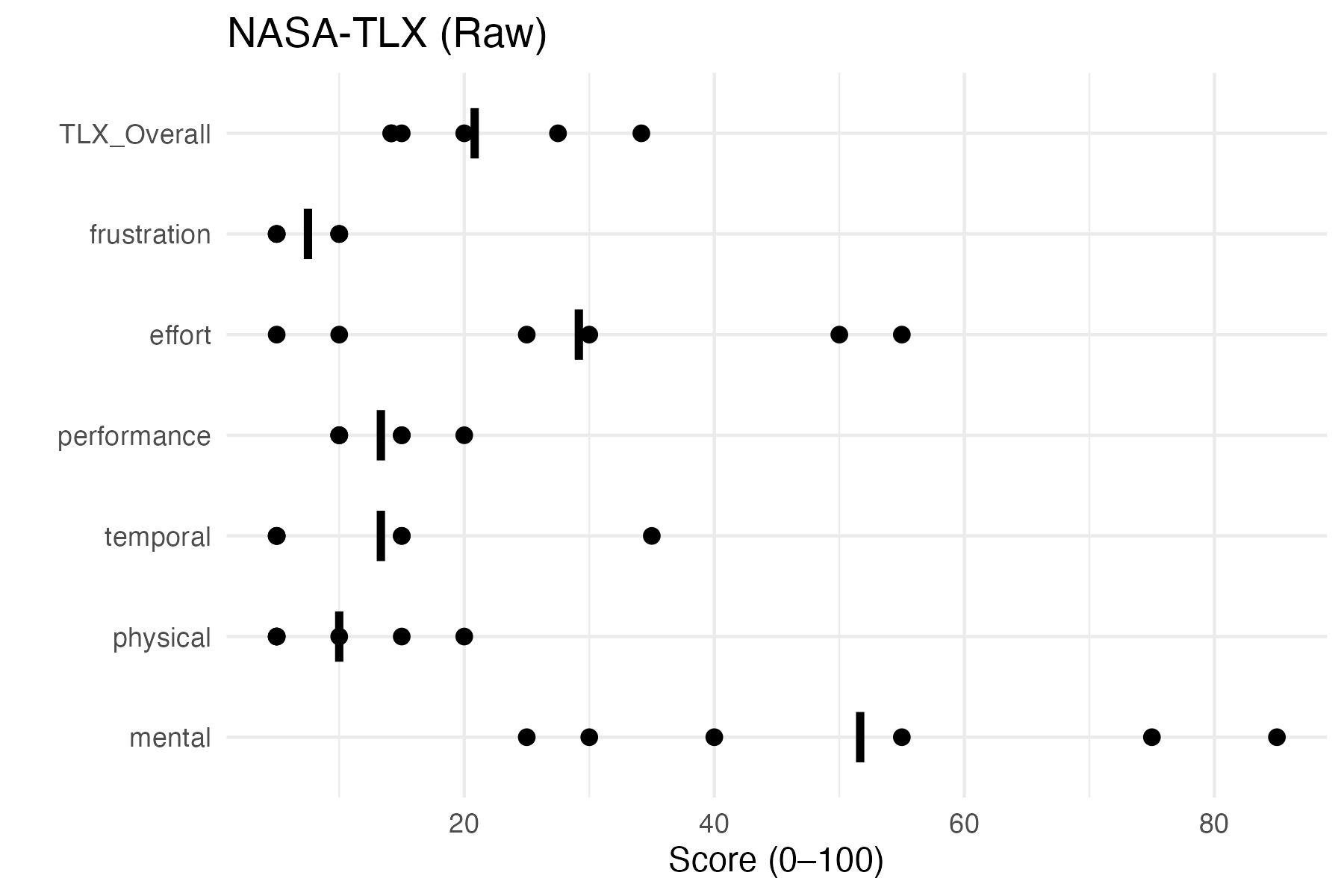}
\caption{NASA-TLX subscales (per-participant points + mean crossbar). Error bars represent 95\% CIs for descriptive purposes only ($N=6$).}
\label{fig:TLX}
\end{figure}

\begin{figure}[h]
\centering
\includegraphics[width=0.85\linewidth]{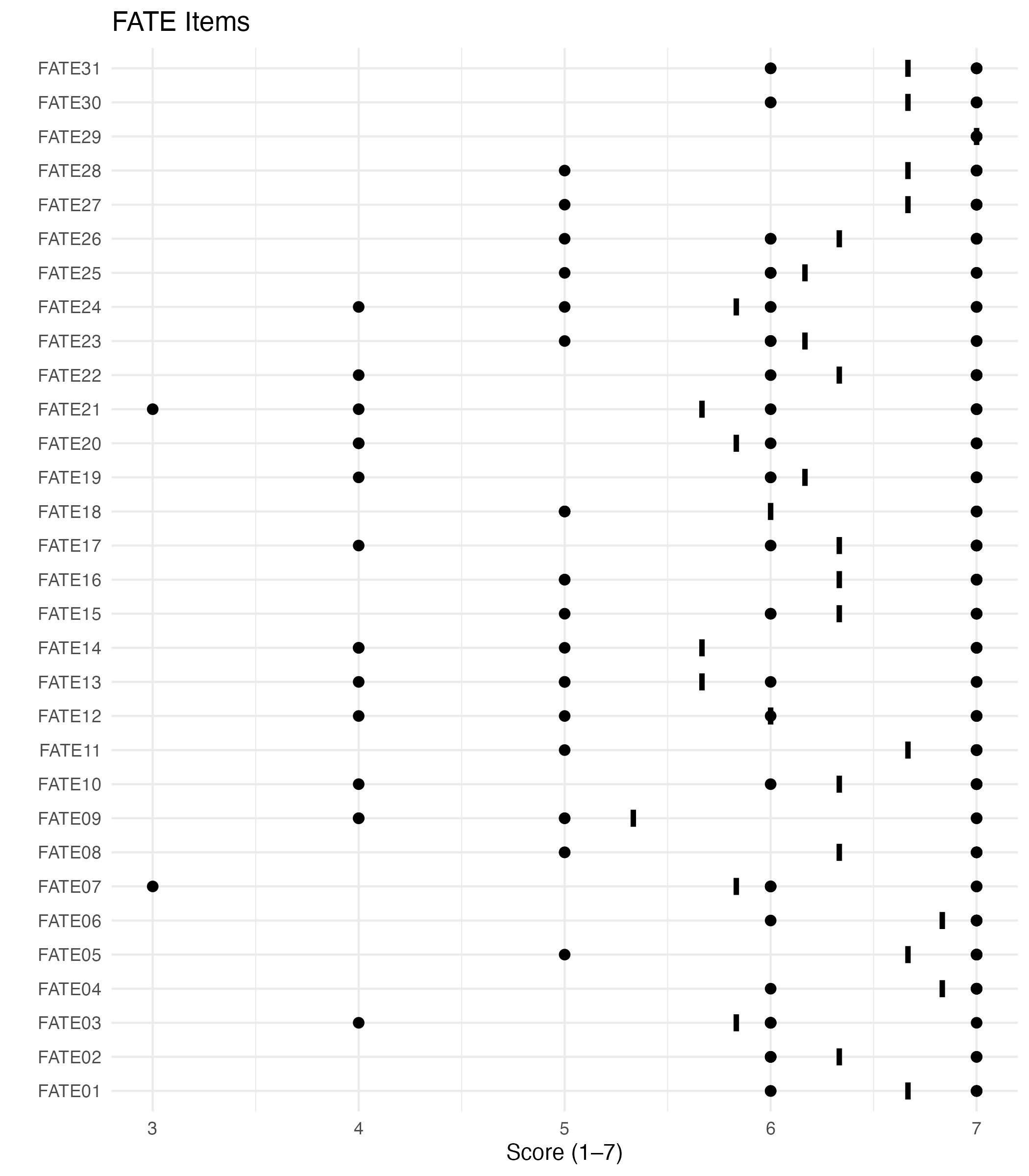}
\caption{FATE constructs and composites (per-participant points + mean crossbar). Error bars represent 95\% CIs for descriptive purposes only ($N=6$).}
\label{fig:FATE}
\end{figure}

\end{document}